\documentclass[aps,prb,amsmath,reprint,superscriptaddress]{revtex4-1}
\usepackage{amsmath,txfonts,bm,color,dcolumn,graphicx,physics}
\begin{document}

\title{Analytical derivative coupling for multistate CASPT2 theory}
\author{Jae Woo Park}
\email{jwpk1201@northwestern.edu}
\author{Toru Shiozaki}
\affiliation{Department of Chemistry, Northwestern University, 2145 Sheridan Rd., Evanston, IL 60208, USA.}
\date{\today}

\begin{abstract}
The probability of non-radiative transitions in photochemical dynamics is determined by the derivative couplings, the couplings between
different electronic states through the nuclear degrees of freedom.
Efficient and accurate evaluation of the derivative couplings is, therefore, of central importance to realize reliable computer simulations of photochemical reactions.
In this work, the derivative couplings for multistate multireference second-order perturbation theory (MS-CASPT2) and 
its `extended' variant (XMS-CASPT2) are studied,
in which we present an algorithm for their analytical evaluation.
The computational costs for evaluating the derivative couplings are essentially the same as those for calculating the nuclear energy gradients.
The geometries and energies calculated with XMS-CASPT2 for small molecules at minimum energy conical intersections (MECIs) are in good agreement with those computed by multireference configuration interaction.
As numerical examples, MECIs are optimized using XMS-CASPT2 for stilbene and a GFP model chromophore (the 4-\textit{para}-hydroxybenzylidene-1,2-dimethyl-imidazolin-5-one anion).
\end{abstract}

\maketitle

\section{Introduction}
Understanding the interaction between molecules and light is an important challenge, not only in basic science but also for technological developments,
because it could lead to the efficient utilization of light in photo-functional materials. 
When molecules are irradiated by photons, the molecules undergo various photochemical processes to relax from their electronic excited states.\cite{Valeurbook}
Non-radiative deactivation is an example of a process that plays a vital role in photo-induced structural changes of molecules
that are used as photochromic and photomechanical materials.\cite{Levine2007ARPC,Erbak-Cakmak2015CR,Matsika2011ARPC,Matsika2014JPCA}
Non-radiative transitions also act as a competitive deactivation pathway in light emission devices,\cite{Olsen2010JACS}
reducing the quantum yield of emission.

These non-radiative transitions are induced by the derivative couplings [also often referred to as nonadiabatic coupling matrix elements (NACMEs)],\cite{Lengsfield1984JCP,Lengsfield1992ACP,Galvan2016JCTC,Lischka2004JCP,Barbatti2008JACS,Tajti2009JCP,Christiansen1999JCP2,Ichino2009JCP,Fatehi2011JCP,Chernyak2000JCP,Send2010JCP,Zhang2014JCP,Zhang2015JCP,Herbert2016ACR}
which are the couplings between the electronic and nuclear degrees of freedom.\footnote{The terms
'derivative coupling' and 'nonadiabatic coupling' are conventionally used interchangeably, except for some recent reports, such as Ref.~\citenum{Galvan2016JCTC}}
We will mathematically define the derivative couplings in the following.
Efficient computation of derivative couplings together with nuclear energy gradients enables
on-the-fly dynamics simulations of photochemical processes.\cite{Tully1990JCP,Levine2007ARPC,Nelson2014ACR,Tavernelli2015ACR}
It also allows for locating conical intersections between potential energy surfaces,\cite{CIbook,Baerbook,Bearpark1994CPL,Manaa1993JCP,Yarkony1996RMP,Matsika2011ARPC}
which are the set of geometries where two or more potential energy surfaces intersect with each other.
Since the computational costs of these applications are strongly dominated by the underlying computation of
the derivative couplings and nuclear energy gradients, development of quantum chemical approaches for their efficient and accurate evaluation
has the potential to significantly advance what is considered to be the state of the art method in computational photophysics and photochemistry.

To achieve efficient evaluation of derivative couplings,
analytical differentiation techniques have been explored in the last few decades.
Historically, analytical derivative couplings for multi-configurational methods were first studied [such as the
state-averaged complete active space self-consistent field (SA-CASSCF)\cite{Lengsfield1984JCP,Lengsfield1992ACP,Galvan2016JCTC}
and uncontracted multireference configuration interaction (unc-MRCI) methods.\cite{Lengsfield1984JCP,Lischka2004JCP}]
These two models nevertheless have disadvantages: SA-CASSCF is essentially a mean-field model with static correlation treatment,
and does not describe dynamical correlation;
unc-MRCI is computational very demanding and is often used without double excitations (i.e., MRCIS that does not describe dynamical correlation).\cite{Barbatti2008JACS}

More recently, the analytical evaluation of derivative couplings for single-reference theories has been extensively investigated,
including those based on equation-of-motion coupled-cluster theory (EOM-CC),\cite{Tajti2009JCP,Christiansen1999JCP2,Ichino2009JCP}
configuration interactions singles (CIS),\cite{Fatehi2011JCP}
time-dependent density functional theories (TDDFT),\cite{Chernyak2000JCP,Send2010JCP}
and their spin-flip variants.\cite{Zhang2014JCP,Zhang2015JCP,Herbert2016ACR}
Standard single-reference methods are, however, known to incorrectly predict the dimensionality of the conical intersection spaces between the ground and excited states,
because they do not compute the states on an equal footing.\cite{Levine2006MP,Huix-Rotllant2013JCTC,Li2014JPCL,Tuna2015JCTC,Herbert2016ACR}
There have been attempts to resolve this problem.\cite{Li2014JPCL,Huix-Rotllant2013JCTC}
In addition, spin-flip single-reference methods that treat the ground and excited states equally have been successfully applied to the computation of the electronic structure around 
conical intersections,\cite{Epifanovsky2007MP,Shepler2008JPCA,Minezawa2009JPCA,Zhang2015JCP,Herbert2016ACR}
yielding conical intersection structures that are in agreement with those obtained with CASSCF or MRCI.
There have nevertheless been discrepancies in the energies of the conical intersections between SF-TDDFT and CASSCF computations.\cite{Herbert2016ACR}
To assess these models, especially for large molecular systems, the development of efficient multi-reference electron correlation methods
for calculating derivative couplings is warranted.

We therefore turn our attention to one of the most successful multi-reference models, complete active space second order perturbation theory (CASPT2).\cite{Andersson1990JPC,Andersson1992JCP}
The CASPT2 method is a post-CASSCF method that describes dynamical correlation up to the second order.
It uses so-called fully internally contracted basis functions to expand the first-order wave functions for efficiency.
The electronic structure around conical intersections can be accurately described by its multistate variant (MS-CASPT2),\cite{Finley1998CPL}
which diagonalizes an effective Hamiltonian formed from the state-specific CASPT2 wave functions.
The MS-CASPT2 method has subsequently been improved by the `extended' variant (XMS-CASPT2)\cite{Shiozaki2011JCP3}
to remedy the erratic behavior of MS-CASPT2 potential energy surfaces when the mixing is strong.\cite{SerranoAndres2005JCP}
Note that this extension was first proposed by Granovsky for uncontracted multireference perturbation theory.\cite{Granovsky2011JCP}
Very recently, One of the authors and co-workers developed an analytical nuclear gradient program for CASPT2,\cite{MacLeod2015JCP,Vlaisavljevich2016JCTC}
which forms the basis for the work presented herein.

In this work, we report the derivation and implementation of the analytical MS-CASPT2 and XMS-CASPT2 derivative couplings.
The computer program has been implemented as an extension of the aforementioned nuclear gradient program.
We note in passing that the interstate couplings studied for MS-CASPT2 (with partial internal contraction\cite{Celani2003JCP,molpro}) by Mori and Kato\cite{Mori2009CPL}
are part of the derivative coupling, which is what we call the MS-mixing term (see below). 
In the following, we first present the definition of the derivative couplings for MS-CASPT2 and XMS-CASPT2 wave functions, followed by 
the working equations for analytically evaluating the (X)MS-CASPT2 derivative couplings.
We compare the minimum energy conical intersections (MECIs) of ethylene optimized by XMS-CASPT2 with those of unc-MRCI.
In addition, the shapes of the potential energy surfaces near the MECIs of a cationic retinal model chromophore (PSB3) are investigated.
We then present the MECIs of stilbene and an anionic GFP model chromophore (\textit{p}-HBDI) optimized by XMS-CASPT2.

\section{Theoretical Background}
In this section, we briefly review the previous work relevant to the development of the analytical (X)MS-CASPT2 derivative couplings.
We first present the XMS-CASPT2 theory that is the basis of this work. 
The definition of the derivative coupling is then presented, followed by an algorithm for the analytical evaluation of the derivative couplings for SA-CASSCF. 

\subsection{XMS-CASPT2 Wave Functions}
XMS-CASPT2 is a quasi-degenerate second-order perturbation theory on the basis of the CASSCF reference functions.
The CASSCF wave functions are a linear combination of the Slater determinants,
\begin{align} \label{CASwfn}
\ket{L} = \sum_I c_{I,L} |I\rangle,
\end{align}
where $c_{I,L}$ are the configuration-interaction (CI) coefficients. 
In the following, $I$ and $J$ label Slater determinants, and $K$, $L$, $M$, and $N$ label reference functions. 
In XMS-CASPT2, the rotated reference functions are formed by diagonalizing the state-averaged Fock operator, $\hat{f}$, within the reference space,\cite{Granovsky2011JCP,Shiozaki2011JCP3,Vlaisavljevich2016JCTC} 
\begin{align}
|\tilde{M}\rangle = \sum_L |L\rangle U_{LM},
\end{align}
where $U_{LM}$ is chosen such that it satisfies
\begin{align}
\sum_{KL}U_{KM} \langle K | \hat{f} | L \rangle  U_{LN} = E^{(0)}_M\delta_{MN}.\label{zeroener}
\end{align}
$E^{(0)}_M$ is the zeroth-order energy of the state $|\tilde{M}\rangle$. For latter convenience, we introduce the rotated reference coefficients, 
\begin{align}
\tilde{c}_{I,M} = \sum_{L} c_{I,L} U_{LM}.
\end{align}
The state-specific CASPT2 wave function is 
the sum of the reference function and the first-order corrections expanded in the internally contracted basis, i.e.,
\begin{align}
|\Phi_M\rangle
= |\tilde{M}\rangle + \sum_{N \Omega} \hat{E}_{\Omega} |\tilde{N}\rangle {T}_{\Omega, LN},
\end{align}
where $\Omega$ denotes all possible double-excitation manifold (see Refs.~\onlinecite{Andersson1990JPC,Andersson1992JCP,MacLeod2015JCP,Vlaisavljevich2016JCTC}).
For brevity, we introduce the following short-hand notations:
\begin{subequations}
\begin{align}
&\hat{T}_{MN} = \sum_\Omega \hat{E}_\Omega T_{\Omega,MN},\\
&|\Phi_M^{(1)}\rangle = \sum_N \hat{T}_{MN} |\tilde{N}\rangle. 
\end{align}
\end{subequations}
The perturbation amplitudes $T_{\Omega,LN}$ are obtained by solving the amplitude equation,
\begin{align}
\langle \tilde{M}| \hat{E}_{\Omega}^\dagger ( \hat{f} - E_N^{(0)} + E_{\text{shift}} ) 
  |\Phi_N^{(1)}\rangle + \langle\tilde{M}| \hat{E}_{\Omega}^\dagger \hat{H} |\tilde{N}\rangle = 0,\label{ampeq}
\end{align}
where $E_\mathrm{shift}$ is the level shift to circumvent intruder state problems.\cite{Roos1995CPL}
Once the amplitudes are determined, the effective
Hamiltonian $H^\text{eff}$ is constructed as
\begin{align}
H^\text{eff}_{LM} = & \langle\tilde{L}| \hat{H} |\tilde{M}\rangle
+ \frac{1}{2} 
\left[ \langle\Phi_{L}^{(1)}| \hat{H} |\tilde{M}\rangle
+      \langle\tilde{L}| \hat{H} |\Phi_M^{(1)}\rangle \right] \nonumber\\
- & \delta_{LM} E_\mathrm{shift}  \langle \Phi_L^{(1)} | \Phi_L^{(1)}\rangle.\label{effhamil}
\end{align}
This effective Hamiltonian is then diagonalized to obtain the XMS-CASPT2 energies and wave functions,
\begin{subequations}
\label{XMSENdef}
\begin{align}
&\sum_M H^{\text{eff}}_{LM} R_{MP} = R_{LP} E_P^{\text{XMS}},\label{heff}\\
&|\Psi_P\rangle = \sum_M |\Phi_M\rangle R_{MP}.
\end{align}
\end{subequations}
Hereafter $P$ and $Q$ label the physical states of interest.
We define the XMS-rotated reference function,
\begin{align}
|P\rangle = \sum_M |\tilde{M}\rangle R_{MP},
\end{align}
which will be used later. 

\subsection{Definition of Derivative Coupling in CASPT2}
The Lagrangian for the total energy of a molecule, including the nuclear and electronic kinetic energy, resulting in the total wave function $\Psi_\mathrm{tot}$ is
\begin{align}
\mathcal{L} = \langle \Psi_\mathrm{tot} | \hat{\mathcal{H}} | \Psi_\mathrm{tot} \rangle - \lambda \left[ \langle {\Psi_\mathrm{tot}} | {\Psi_\mathrm{tot}} \rangle - 1 \right], \label{totlag}
\end{align}
and solving the stationary condition for this Lagrangian, $\partial \mathcal{L} / \partial \Psi_\mathrm{tot} = 0$, leads to the molecular Schr\"odinger equation.
Here, the molecular Hamiltonian $\hat{\mathcal{H}}$ is written as a sum of the nuclear kinetic energy operator $\hat{T}_\mathrm{nuc}$
and the electronic Hamiltonian $\hat{H}_\mathrm{el}$,
\begin{align}
\hat{\mathcal{H}} \left( \mathbf{r}, \mathbf{X} \right) = \hat{T}_\mathrm{nuc} \left(\mathbf{X} \right) + \hat{H}_\mathrm{el} \left( \mathbf{r}, \mathbf{X} \right),
\end{align}
where $\mathbf{r}$ and $\mathbf{X}$ are the electronic and nuclear coordinates. The nuclear kinetic energy operator is explicitly written as
\begin{align}
\hat{T}_\mathrm{nuc} = -\sum_A \frac{1}{2 M_A} \frac{d^2}{d \mathbf{X}_A^2},
\end{align}
where $A$ labels the nuclei, and $M_A$ is the nuclear mass.
The expectation value for the total energy $\mathcal{E}$ with $\Psi_\mathrm{tot}$ is
\begin{align}
\mathcal{E} = \langle \Psi_\mathrm{tot} | \hat{\mathcal{H}} | \Psi_\mathrm{tot} \rangle. \label{toten}
\end{align}
Within the Born--Oppenheimer approximation, one ignores the $\hat{T}_\mathrm{nuc}$
term by assuming that the kinetic energies of the nuclei are much smaller than the kinetic energies of the electrons. Then, the Lagrangian for the 
electronic Hamiltonian becomes
\begin{align}
\mathcal{L}_{\mathrm{BO},\mathbf{X}} &= 
\langle \Psi_\mathrm{el} (\mathbf{r}; \mathbf{X}) | \hat{H}_\mathrm{el} | \Psi_\mathrm{el} (\mathbf{r}; \mathbf{X}) \rangle \nonumber \\
 & - \lambda \left[ \langle {\Psi_\mathrm{el} (\mathbf{r}; \mathbf{X})} | {\Psi_\mathrm{el} (\mathbf{r}; \mathbf{X})} \rangle - 1 \right],
\end{align}
where $| \Psi_\mathrm{el} \rangle$ is an electronic wave function. The stationary condition for this Lagrangian, 
\begin{align}
\frac {\partial \mathcal{L}_{\mathrm{BO},\mathbf{X}}}{\partial \Psi_\mathrm{el} (\mathbf{r}; \mathbf{X})} = 0,
\end{align}
leads to the electronic Schr\"odinger equation.
The set of the solutions for this equation yields the adiabatic basis set (for example, the full CI eigenvectors and their corresponding energies). We will
denote the individual adiabatic state as $| \Psi_{\mathrm{el},P} (\mathbf{r}; \mathbf{X}) \rangle$.
To calculate the total energy with the motions of the electrons and nuclei using the electronic wave functions, we expand the total wave function in the adiabatic basis set
\begin{align}
| \Psi_\mathrm{tot} (\mathbf{r}, \mathbf{X}) \rangle = \sum_P | \Psi_{\mathrm{el},P} (\mathbf{r}; \mathbf{X}) \rangle \chi_P (\mathbf{X}), \label{adiabexp}
\end{align}
where the expansion coefficients $\chi$ are normalized. Substituting this into the expression for the total energy $\mathcal{E}$ [Eq.~\eqref{toten}] yields the expression
\begin{align}
\mathcal{E} = \sum_P \chi_P^2 \left( T_{\mathrm{nuc},P} + E_P \right) + \sum_{Q \neq P} \chi_Q \langle \Psi_{\mathrm{el},Q} | \hat{T}_\mathrm{nuc} | \Psi_{\mathrm{el},P} \rangle \chi_P,\label{etot}
\end{align}
The off-diagonal elements of the nuclear kinetic energy term can be explicitly written as
\begin{align}
\langle \Psi_{\mathrm{el},Q} | \hat{T}_\mathrm{nuc} | \Psi_{\mathrm{el},P} \rangle
&= -\sum_{A} \frac{1}{M_A} \braket{ \Psi_{\mathrm{el},Q} }{ \frac{d\Psi_{\mathrm{el},P}}{d\mathbf{X}_A} } \cdot \frac{d}{d\mathbf{X}_A}\nonumber\\
&  -\frac{1}{2} \sum_{A} \frac{1}{M_A}\braket{ \Psi_{\mathrm{el},Q} }{\frac{d^2\Psi_{\mathrm{el},P}}{d\mathbf{X}_A^2}} ,
\label{couplings}
\end{align}
where $A$ labels the nuclei, and $M_A$ is the nuclear mass. The matrix element in the first term is the derivative coupling between the adiabatic states $Q$ and $P$,
which we hereafter denote as $\mathbf{d}^{QP}$,
\begin{align}
\mathbf{d}^{QP} = \braket{\Psi_{\mathrm{el},Q}}{\frac{d\Psi_{\mathrm{el},P}}{d \mathbf{X}} }. \label{defdc}
\end{align}
The second term on the right hand side of Eq.~\eqref{couplings} is the scalar coupling.\cite{Lengsfield1984JCP,Tully1990JCP,Lengsfield1992ACP,CIbook,Baerbook}

The derivative coupling formulas for (X)MS-CASPT2 can be obtained in a similar fashion. The Lagrangians for the zeroth-order energy and the energy corrected to second order (the Hylleraas functional) are
\begin{subequations}
\begin{align}
&\mathcal{L}^{(0)} = \langle \Psi_\mathrm{tot}^{(0)} | \hat{\mathcal{H}} | \Psi_\mathrm{tot}^{(0)} \rangle - \lambda \left[ \langle {\Psi_\mathrm{tot}^{(0)}} | {\Psi_\mathrm{tot}^{(0)}} \rangle - 1 \right], \\
&\mathcal{L}^{(2)} = \langle \Psi^{(1)}_\mathrm{tot} | \hat{\mathcal{H}}^{(0)} - \mathcal{E}^{(0)} | \Psi_\mathrm{tot}^{(1)} \rangle + 2 \langle \Psi^{(1)}_\mathrm{tot} | \hat{\mathcal{H}} | \Psi^{(0)}_\mathrm{tot} \rangle,\label{totalhy}
\end{align}
\end{subequations}
where the zeroth-order Hamiltonian and the zeroth-order total energy are defined as
\begin{subequations}
\begin{align}
&\hat{\mathcal{H}}^{(0)} = \hat{T}_\mathrm{nuc} + \hat{H}_\mathrm{el}^{(0)},\\
&\mathcal{E}^{(0)} = \langle \Psi^{(0)}_\mathrm{tot} | \hat{\mathcal{H}}^{(0)} | \Psi^{(0)}_\mathrm{tot} \rangle.
\end{align}
\end{subequations}
$\hat{H}_\mathrm{el}^{(0)}$ is the zeroth-order Hamiltonian in XMS-CASPT2. The expressions for the
energy corrected to second order $\mathcal{E}^\mathrm{PT2}$ with optimized $\Psi_\mathrm{tot}^{(0)}$ and $\Psi_\mathrm{tot}^{(1)}$ are
\begin{align}
\label{toten2}
&\mathcal{E}^\mathrm{PT2} = \langle \Psi_\mathrm{tot}^{(0)} | \hat{\mathcal{H}} | \Psi_\mathrm{tot}^{(0)} \rangle + \langle \Psi_\mathrm{tot}^{(1)} | \hat{\mathcal{H}} | \Psi_\mathrm{tot}^{(0)} \rangle.
\end{align}
Again, within the Born--Oppenheimer approximation, one ignores the $\hat{T}_\mathrm{nuc}$ term, and the Lagrangians become
\begin{subequations}
\begin{align}
\mathcal{L}^{(0)}_{\mathrm{BO},\mathbf{X}} &= 
\langle \Psi^{(0)}_\mathrm{el} (\mathbf{r}; \mathbf{X}) | \hat{H}_\mathrm{el} | \Psi^{(0)}_\mathrm{el} (\mathbf{r}; \mathbf{X}) \rangle \nonumber \\
 & - \lambda \left[ \langle {\Psi_\mathrm{el}^{(0)} (\mathbf{r}; \mathbf{X})} | {\Psi_\mathrm{el}^{(0)} (\mathbf{r}; \mathbf{X})} \rangle - 1 \right], \\
\mathcal{L}^{(2)}_{\mathrm{BO},\mathbf{X}} & = \langle \Psi_\mathrm{el}^{(1)} (\mathbf{r}; \mathbf{X}) | \hat{H}_\mathrm{el}^{(0)} - E_\mathrm{el}^{(0)} | \Psi_\mathrm{el}^{(1)} (\mathbf{r}; \mathbf{X}) \rangle \nonumber \\
& + 2 \langle \Psi_\mathrm{el}^{(1)} (\mathbf{r}; \mathbf{X}) | \hat{H}_\mathrm{el} | \Psi_\mathrm{el}^{(0)} (\mathbf{r}; \mathbf{X}) \rangle,
\end{align}
\end{subequations}
where $E_\mathrm{el}^{(0)}$ is the zeroth-order electronic energy [see Eq.~\eqref{zeroener}]. The stationary conditions for $\mathcal{L}^{(0)}_{\mathrm{BO},\mathbf{X}}$, and
$\mathcal{L}^{(1)}_{\mathrm{BO},\mathbf{X}}$ with fixed $\Psi_{\mathrm{el}}^{(0)}(\mathbf{r};\mathbf{X})$,
\begin{subequations}
\begin{align}
& \frac {\partial \mathcal{L}^{(0)}_{\mathrm{BO},\mathbf{X}}}{\partial \Psi_\mathrm{el}^{(0)} (\mathbf{r}; \mathbf{X})} = 0,\\
& \left. \frac {\partial \mathcal{L}^{(2)}_{\mathrm{BO},\mathbf{X}}}{\partial \Psi_\mathrm{el}^{(1)} (\mathbf{r}; \mathbf{X})} \right|_{\Psi_{\mathrm{el}}^{(0)} (\mathbf{r}; \mathbf{X})} = 0,
\end{align}
\end{subequations}
give us a set of the zeroth- and first-order wave functions [the latter corresponds to Eq.~\eqref{ampeq}].
Analogous to Eq.~\eqref{adiabexp}, we expand
the total wave functions in the basis of the electronic wave functions as
\begin{subequations}
\begin{align}
| \Psi_\mathrm{tot}^{(0)} (\mathbf{r}, \mathbf{X})\rangle &= \sum_M | \Psi_{\mathrm{el},M}^{(0)} (\mathbf{r}; \mathbf{X}) \rangle \chi_M (\mathbf{X}), \\
| \Psi_\mathrm{tot}^{(1)} (\mathbf{r}, \mathbf{X})\rangle &= \sum_M | \Psi_{\mathrm{el},M}^{(1)} (\mathbf{r}; \mathbf{X}) \rangle \chi_M (\mathbf{X}),
\end{align}
\end{subequations}
By substituting these into the total second-order energy $\mathcal{E}^\mathrm{PT2}$ [Eq.~\eqref{toten2}], we obtain
\begin{align}
\mathcal{E}^\mathrm{PT2} 
  &= \sum_P \chi_P^{\prime 2} \left( T_{\mathrm{nuc},P} + E_P^\mathrm{XMS} \right)  + \sum_{Q \neq P} \chi_Q^{\prime} \langle Q | \hat{T}_\mathrm{nuc} | \Psi_P \rangle \chi_P^{\prime} \nonumber \\
  &= \sum_P \chi_P^{\prime 2} \left( T_{\mathrm{nuc},P} + E_P^\mathrm{XMS} \right) \nonumber \\
  &+ \frac{1}{2} \sum_{Q \neq P} \chi_Q^{\prime} \left( \langle Q | \hat{T}_\mathrm{nuc} | \Psi_P \rangle + \langle \Psi_Q | \hat{T}_\mathrm{nuc} | P \rangle \right)\chi_P^{\prime}.
\label{subst}
\end{align}
Note that this substitution is an approximation if the total wave functions are not fully optimized.
In Eq.~\eqref{subst}, the electronic wave functions are rotated to make $\hat{H}_\mathrm{el}$ diagonal [which is Eq.~\eqref{heff}],
\begin{subequations}
\begin{align}
&| P \rangle =  | \Psi_{\mathrm{el},M}^{(0)} \rangle R_{MP}, \\
&| \Psi_P \rangle =  \left(  | \Psi_{\mathrm{el},M}^{(0)} \rangle + | \Psi_{\mathrm{el},M}^{(1)} \rangle \right) R_{MP}, \\
&\chi_P^{\prime} =  \chi_M R_{MP}.
\end{align}
\end{subequations}
The off-diagonal elements of the $\hat{T}_\mathrm{nuc}$ term can therefore be explicitly written as
\begin{align}
&\frac{1}{2} \left( \langle Q | \hat{T}_\mathrm{nuc} | \Psi_P \rangle + \langle \Psi_Q | \hat{T}_\mathrm{nuc} | P \rangle \right)\nonumber\\
&\quad =-\frac{1}{2}\sum_A \frac{1}{M_A} \left[ \braket{ Q }{ \frac{d\Psi_P}{d\mathbf{X}_A} } + \braket{ \Psi_Q }{ \frac{dP}{d\mathbf{X}_A} } \right] \cdot \frac{d}{d\mathbf{X}_A}  \nonumber \\
&\quad -\frac{1}{4}\sum_A \frac{1}{M_A}\left[ \braket{ Q }{\frac{d^2\Psi_P}{d\mathbf{X}_A^2}}  + \braket{ \Psi_Q }{\frac{d^2P}{d\mathbf{X}_A^2} } \right],
\label{couplingspt2}
\end{align}
which leads to the following expression for the derivative coupling in (X)MS-CASPT2,
\begin{align}
\mathbf{d}^{\mathrm{PT2},QP} = \frac{1}{2} \left[ \braket{Q}{\frac{d \Psi_P}{d \mathbf{X}}} + \braket{\Psi_Q}{\frac{d P}{d \mathbf{X}}} \right].\label{NEdef}
\end{align}
The validity of the expression for $\mathbf{d}^{\mathrm{XMS},QP}$ has been confirmed by calculating the line integral around conical intersections (see below).\cite{CIbook,Baerbook}
This quantity can be readily used in many formulations of non-adiabatic dynamics,
for instance in fewest-switch surface-hopping (FSSH) non-adiabatic dynamics.\cite{Tully1990JCP,Fabiano2008CP}
Using the derivative coupling defined in Eq.~\eqref{NEdef}, the equation for XMS-CASPT2 FSSH dynamics that propagates $\chi_Q$ is 
\begin{align}
i \frac{\partial\chi_Q}{\partial t} = \chi_Q E^\mathrm{PT2}_Q - i \sum_P \chi_P \mathbf{v}\cdot \mathbf{d}^{\mathrm{PT2},QP},
\end{align}
where we introduce the velocity of the trajectory as $\mathbf{v} = d\mathbf{X}/dt$.

\subsection{Analytical SA-CASSCF Derivative Coupling}
The derivative coupling between SA-CASSCF wave functions
can be written as
\begin{subequations}
\label{casdc}
\begin{align}
&\mathbf{d}^{QP} = \mathbf{d}^{QP}_\mathrm{CI} + \mathbf{d}^{QP}_\mathrm{det},\\ 
&\mathbf{d}^{QP}_\mathrm{CI} =  \sum_I c_{I,Q} \frac{d c_{I,P}}{d \mathbf{X}},\\
&\mathbf{d}^{QP}_\mathrm{det} = \sum_{IJ} c_{I,Q} c_{J,P} \braket{I}{\frac{d J}{d \mathbf{X}}}.
\end{align}
\end{subequations}
We call $\mathbf{d}^{QP}_\mathrm{CI}$ and $\mathbf{d}^{QP}_\mathrm{det}$ the CI and determinant terms, respectively.\cite{Lischka2004JCP,Galvan2016JCTC,Lengsfield1984JCP,CIbook}
The CI term can be evaluated using the following relationship,
\begin{align}
\mathbf{d}^{QP}_\mathrm{CI}
= \frac{1}{E^\mathrm{CAS}_P - E^\mathrm{CAS}_Q}\sum_{IJ} c_{I,Q} \frac{d H_{IJ}} {d \mathbf{X}} c_{J,P}, \label{citerm_casscf}
\end{align}
where $E^\mathrm{CAS}_P$ and $E^\mathrm{CAS}_Q$ are the CASSCF energies of states $P$ and $Q$.
Note that this term corresponds to the so-called interstate couplings,\cite{CIbook,Bearpark1994CPL,Manaa1993JCP,Mori2009CPL,Matsika2011ARPC}
and are analogous to the Hellmann--Feynman forces in nuclear gradient theory.\cite{CIbook,Lengsfield1992ACP,Matsika2011ARPC}
This formula follows from the fact that the off-diagonal elements of the
Hamiltonian in the basis of the SA-CASSCF states are zero, $\matrixel{Q}{\hat{H}}{P} = 0$.
Since the SA-CASSCF wave functions are optimized with respect to
both the CI and orbital coefficients, we can write
the Lagrangian\cite{Celani2003JCP,Koch1990JCP2} for the CI term multiplied by $E_{P}^\mathrm{CAS} - E_{Q}^\mathrm{CAS}$ as 
\begin{align}
\mathcal{L}^{QP} &= \mathbf{c}_Q^\dagger \mathbf{H} \mathbf{c}_P
+ \frac{1}{2} \text{tr} \left[ \mathbf{Z}^{\dagger} ( \mathbf{A} - \mathbf{A}^\dagger ) \right]
- \frac{1}{2} \text{tr} \left[ \mathbf{V} ( \mathbf{C}^{\dagger} \mathbf{SC} - \mathbf{1} ) \right] \nonumber\\
&+ \sum_{N} W_{N} \left[ \mathbf{z}_N^{\dagger} ( \mathbf{H} - E^\mathrm{CAS}_{N} ) \mathbf{c}_{N}
- \frac{1}{2}x_{N} ( \mathbf{c}_N^{\dagger} \mathbf{c}_N - 1 ) \right].
\label{caslag}
\end{align}
The first term is the off-diagonal element of the CI Hamiltonian, and the remaining terms represent the SA-CASSCF convergence conditions, where
$\mathbf{Z}$, $\mathbf{V}$, $\mathbf{z}_N$, and $x_N$ are their Lagrange multipliers.
$\mathbf{A}$ is the derivative of the SA-CASSCF energy with respect to the orbital rotations,
$\mathbf{C}$ and $\mathbf{S}$ are the orbital coefficients and overlap matrix elements, and  
$W_N$ and $\mathbf{c}_N$ are the weight and CI coefficients of state $N$.
See details in Refs.~\onlinecite{Shiozaki2011JCP3} and \onlinecite{Celani2003JCP}.
The so-called $Z$-vector equation\cite{Handy1984JCP,Lengsfield1992ACP} is obtained by differentiating the Lagrangian
with respect to the orbital rotation parameters and CI coefficients;
the source terms of the $Z$-vector equation are
\begin{subequations}
\begin{align}
&Y_{rs} =  \frac{\partial \mathcal{L}^{QP}} {\partial \kappa_{rs}}, \\ 
&y_{I,M} =  \frac{\partial \mathcal{L}^{QP}} {\partial c_{I,M}} = 0.
\end{align}
\end{subequations}
The multiplier-dependent terms in the $Z$-vector equation are identical to those in nuclear gradient theory.\cite{Celani2003JCP,Shiozaki2011JCP3,Gyorffy2013JCP}
After solving the $Z$-vector equation for $\mathbf{Z}$ and $\mathbf{z}$,
we compute the effective density matrices ($\gamma^\mathrm{eff}$ and $\Gamma^\mathrm{eff}$)
and the Lagrange multipliers $\mathbf{V}$, whose explicit forms are given in Refs.~\onlinecite{Shiozaki2011JCP3} and \onlinecite{Celani2003JCP}.
These matrices are then contracted with the derivative integrals to yield the CI term:
\begin{align}
\label{CIterm}
&\mathbf{d}^{QP}_\mathrm{CI} = \frac{1}{E^\mathrm{CAS}_P - E^\mathrm{CAS}_Q}\nonumber\\
&\quad \times \left[\sum_{\mu\nu} h_{\mu\nu}^\mathbf{X} \gamma^{\text{eff}}_{\mu\nu} + 
  \sum_{\mu\nu\lambda\sigma} \left( \mu\nu | \lambda\sigma \right)^\mathbf{X} \Gamma^{\text{eff}}_{\mu\nu\lambda\sigma} + 
  \sum_{\mu\nu} S_{\mu\nu}^\mathbf{X} V_{\mu\nu}\right],
\end{align}
where $\mu$, $\nu$, $\lambda$ and $\sigma$ label atomic orbitals, and the superscript ${\mathbf{X}}$ denotes the integral derivatives with respect to $\mathbf{X}$.
The formula for the determinant term can be easily derived if one rewrites the operator $d / d \mathbf{X}$
as a one-electron operator,\cite{Lengsfield1984JCP}
\begin{align}
&\frac{d}{d \mathbf{X}} = \sum_{rs} \left( \frac{d \kappa_{rs}}{d \mathbf{X}} + \sigma_{rs}^\mathbf{X} \right) \hat{E}_{rs},\\
&\sigma_{rs}^\mathbf{X} = \sum_{\mu\nu} C_{\mu r}C_{\nu s} \braket{\phi_\mu}{\frac{d \phi_\nu}{d \mathbf{X}}}.
\end{align}
in which $\phi_\mu$ are atomic orbitals, $d \kappa_{rs} / d \mathbf{X}$ are the orbital response parameters, and $\hat{E}_{rs}$ is the spin-adapted one-particle excitation operator.
This leads to a compact form of the determinant term,
\begin{align}
\mathbf{d}^{QP}_\mathrm{det}
= \sum_{rs} \gamma_{rs}^{QP} \left(\frac{d\kappa_{rs}}{d\mathbf{X}} + \sigma_{rs}^\mathbf{X}\right),
\label{detterm}
\end{align}
where $\gamma^{QP}_{rs} = \langle Q| \hat{E}_{rs}|P\rangle$.
In practice, the evaluation of the derivative of $\kappa_{rs}$ can be avoided in one of two ways:
One approach is to include it in the $Z$-vector algorithm by adding the following $Y^C_{rs}$ to $Y_{rs}$,
\begin{align}
Y^C_{rs} = (E^\mathrm{CAS}_P - E^\mathrm{CAS}_Q) \gamma_{rs}^{QP}.
\end{align}
The other approach is to use the following equivalent expression\cite{Lischka2004JCP,Galvan2016JCTC} 
that is written in terms of $\sigma_{rs}^{X}$ alone,
\begin{align}
\mathbf{d}^{QP}_\mathrm{det}
= \frac{1}{2} \sum_{rs} ( \gamma_{rs}^{QP} - \gamma_{sr}^{QP} ) \sigma_{rs}^\mathbf{X}.
\label{equiva}
\end{align}
The latter approach does not require modification of the $Z$-vector equation.
A similar algorithm to the former is used in the evaluation of the XMS-CASPT2 derivative couplings described below.

\section{Analytical XMS-CASPT2 Derivative Coupling}
The XMS-CASPT2 derivative coupling can be formally written as 
\begin{subequations}
\label{XMSExplicit}
\begin{align}
& \mathbf{d}^{\mathrm{XMS},QP}  =  \mathbf{d}^{\mathrm{XMS},QP}_\mathrm{mix} + \mathbf{d}^{\mathrm{XMS},QP}_\mathrm{CAS} + \mathbf{d}^{\mathrm{XMS},QP}_\mathrm{PT2},\\ 
& \mathbf{d}^{\mathrm{XMS},QP}_\mathrm{mix} = \sum_K R_{KQ} \frac{d R_{KP}}{d \mathbf{X}}, \\
& \mathbf{d}^{\mathrm{XMS},QP}_\mathrm{CAS} = \sum_{KL} \mathcal{R}^{QP}_{KL}  
\braket{\tilde{K}}{\frac{d \tilde{L}}{d \mathbf{X}}}, \\ 
& \mathbf{d}^{\mathrm{XMS},QP}_\mathrm{PT2} = \sum_{KL} \mathcal{R}^{QP}_{KL}  
\braket{\Phi_K^{(1)}}{\frac{d \tilde{L}}{d \mathbf{X}}},
\end{align}
\end{subequations}
where we introduced a short-hand notation for the products of the MS mixing matrices, 
\begin{align}
\mathcal{R}^{QP}_{KL} = \frac{1}{2}(R_{KQ} R_{LP} - R_{KP} R_{LQ}). 
\end{align}
The expressions in Eq.~\eqref{XMSExplicit} are manifestly invariant with respect to any rotations among active orbitals.
The orbital invariance has also been confirmed numerically.
In the following, we present the working equations for its analytical evaluation.

Since the XMS-CASPT2 states are obtained from the eigenvalue equation, Eq.~\eqref{XMSENdef},
the MS-mixing term $\mathbf{d}^{\mathrm{XMS},QP}_\mathrm{mix}$ can be evaluated using a strategy similar to the one for the CI term [Eq.~\eqref{citerm_casscf}] in the evaluation of the SA-CASSCF derivative coupling,
\begin{align}\label{PT2mix}
\mathbf{d}^{\mathrm{XMS},QP}_\mathrm{mix} = 
\frac{1}{\Delta E_{PQ}^\mathrm{XMS}}\sum_{KL} R_{KQ} \frac{d H_{KL}^{\text{eff}}} {d \mathbf{X}} R_{LP},
\end{align}
in which we define $\Delta E_{PQ}^\mathrm{XMS} = E_P^{\text{XMS}} - E_Q^{\text{XMS}}$ for brevity.
This MS-mixing term corresponds to the interstate coupling
and can be evaluated using the methodologies described in previous work\cite{Mori2009CPL,Vlaisavljevich2016JCTC}
using the PT2 Lagrangian,
\begin{align}
\mathcal{L}_{\text{PT2}}^{QP} &= \sum_{KL} R_{KQ} H^{\text{eff}}_{KL} R_{LP} \nonumber\\ 
&+ \sum_{KMN} \langle\tilde{M}|\hat{\lambda}^{\dagger}_{KM} ( \hat{f} - E_K^{(0)} + E_\text{shift} ) \hat{T}_{KN}|\tilde{N}\rangle \nonumber\\
&+ \sum_{KM}  \langle\tilde{M}|\hat{\lambda}^{\dagger}_{KM} \hat{H}|\tilde{K}\rangle.
\end{align}
The so-called $\lambda$-equation is obtained by making the PT2 Lagrangian stationary with respect to the perturbation amplitudes,
$T_{\Omega,KN}$:
\begin{align}
\frac{\partial \mathcal{L}^{QP}_{\text{PT2}}} {\partial T_{\Omega,KN}} = 0.
\end{align}
The explicit expression for the total Lagrangian is
\begin{align}
\label{totallag}
\mathcal{L}^{QP} &= \mathcal{L}^{QP}_{\text{PT2}} 
+ \frac{1}{2} \text{tr} \left[ \mathbf{Z}^{\dagger} ( \mathbf{A} - \mathbf{A}^\dagger ) \right]
- \frac{1}{2} \text{tr} \left[ \mathbf{V} ( \mathbf{C}^{\dagger} \mathbf{SC} - \mathbf{1} ) \right] \nonumber\\
&+ \sum_{N} W_{N} \left[ \mathbf{z}_N^{\dagger} ( \mathbf{H} - E^\mathrm{CAS}_{N} ) \mathbf{c}_{N}
- \frac{1}{2}x_{N} ( \mathbf{c}_N^{\dagger} \mathbf{c}_N - 1 ) \right] \nonumber\\
&+ \sum_i^\mathrm{closed} \sum_j^\mathrm{frozen} z_{ij}^c f_{ij} + \sum_{MN} w_{MN} f_{MN}. 
\end{align}
The second, third, and fourth terms on the right-hand side account for the SA-CASSCF stationary conditions and also appeared in Eq.~\eqref{caslag}.
The last two terms account for the frozen-core approximation and the XMS rotation, respectively.
Using this Lagrangian, the MS-mixing term is simply written as
\begin{align}
\mathbf{d}^{\mathrm{XMS},QP}_\mathrm{mix} = 
\frac{1}{\Delta E_{PQ}^\mathrm{XMS}} \left(\frac{d \mathcal{L}^{QP}}{d \mathbf{X}}\right)_R,
\label{xmst}
\end{align}
where the subscript $R$ indicates that the MS rotation matrix elements in the first term of the Lagrangian are held fixed when taking the derivative. 
The CAS and PT2 terms are
\begin{subequations}
\label{cast}
\begin{align}
\mathbf{d}^{\mathrm{XMS},QP}_\mathrm{CAS} &= \sum_{KLM} \mathcal{R}^{QP}_{KL} U_{MK} \frac{d U_{ML}}{d \mathbf{X}} 
\nonumber\\ &+ \sum_{KLMN} \mathcal{R}^{QP}_{KL} U_{MK} U_{NL} \sum_I c_{I,M} \frac{d c_{I,N}}{d \mathbf{X}} \nonumber\\
& + \sum_{KL} \mathcal{R}^{QP}_{KL} \sum_{rs} \tilde{\gamma}_{rs}^{KL} \left(\frac{d\kappa_{rs}}{d\mathbf{X}} + \sigma_{rs}^{\mathbf{X}}\right)\\
\mathbf{d}_\mathrm{PT2}^{\mathrm{XMS},QP} &= 
\sum_{KL} \mathcal{R}^{QP}_{KL}   
\sum_{rs}\langle \Phi^{(1)}_K|\hat{E}_{rs}|\tilde{L}\rangle \left(\frac{d \kappa_{rs}}{d\mathbf{X}} + \sigma_{rs}^{\mathbf{X}}\right).
\end{align}
\end{subequations}
To avoid the evaluation of the derivatives of $U_{MK}$, $c_{I,N}$, and $\kappa_{rs}$ with respect to $\mathbf{X}$, we simultaneously evaluate Eqs.~\eqref{xmst} and \eqref{cast} 
using the $Z$-vector equation.
First, the Lagrange multiplier $w_{MN}$ is evaluated as
\begin{align}
w_{MN} &= \frac{1}{E^{(0)}_N - E^{(0)}_M} \left[ \frac{1}{2}\sum_{I} \left(\tilde{c}_{I,M} \frac{\partial \mathcal{L}_{\text{PT2}}^{QP}}{\partial \tilde{c}_{I,N}} - \tilde{c}_{I,N}\frac{\partial \mathcal{L}_{\text{PT2}}^{QP}}{\partial \tilde{c}_{I,M}}\right)\right.\nonumber\\
&+ \Delta E_{PQ}^\mathrm{XMS} \sum_{KL} \mathcal{R}^{QP}_{KL} U_{MK} U_{NL} \Bigg].
\end{align}
The $Z$-vector algorithm to calculate  $\mathbf{Z}$, $\mathbf{z}_N$, $\mathbf{V}$, and $x_N$
is analogous to the one used in the nuclear gradient algorithm;\cite{Celani2003JCP,Shiozaki2011JCP3,Vlaisavljevich2016JCTC}
the source terms for the $Z$-vector equation are
\begin{subequations}
\begin{align}
\label{largey}
&Y_{rs} = \frac{\partial \mathcal{L}_{\text{PT2}}^{QP}}{\partial \kappa_{rs}} + \Delta E_{PQ}^\mathrm{XMS}\sum_{KL} \mathcal{R}^{QP}_{KL} \left[\tilde{\gamma}_{rs}^{KL} +
\langle \Phi^{(1)}_K|\hat{E}_{rs}|\tilde{L}\rangle \right] ,\\ 
&y_{I,N} = \sum_M \left[\frac{\partial \mathcal{L}_{\text{PT2}}^{QP}} {\partial \tilde{c}_{I,M}} + \Delta E_{PQ}^\mathrm{XMS} \mathcal{R}^{QP}_{LM} \tilde{c}_{I,L} \right]U_{NM}.
\label{smally}
\end{align}
\end{subequations}
The contribution from the second term in the square bracket in Eq.~\eqref{smally} vanishes in the case of XMS-CASPT2 (but not in MS-CASPT2),
because $y_{I,N}$ is taken to be orthogonal to the reference space when XMS reference functions are used.
The $Z$-vector equation is solved iteratively.
Having determined $w_{MN}$, $\mathbf{Z}$, $\mathbf{z}_N$, and $\mathbf{V}$, the derivative couplings $\mathbf{d}^{\mathrm{XMS},QP}$ can be computed as 
\begin{align}
\mathbf{d}^{\mathrm{XMS},QP} &= \frac{1}{\Delta E_{PQ}^\mathrm{XMS}}\left(\frac{\partial \mathcal{L}^{QP}}{\partial \mathbf{X}}\right)_R \nonumber\\
&+ \sum_{rs}\sum_{KL}  \mathcal{R}^{QP}_{KL} \left[\tilde{\gamma}_{rs}^{KL} + \langle \Phi^{(1)}_K|\hat{E}_{rs}|\tilde{L}\rangle \right] \sigma^\mathbf{X}_{rs}.
\end{align}
The first term is computed as a contraction of the effective density matrices with the derivative integrals as in the nuclear gradient algorithms. 
The algorithm for MS-CASPT2 can be obtained by setting $U_{MN} = \delta_{MN}$ and neglecting its derivative.
If desired, translational invariance can be achieved by setting the second term to zero.\cite{Fatehi2011JCP,Fatehi2012JPCL}
Note that density fitting is used when evaluating the above expressions in our implementation.
The additional computational cost associated with the CAS and PT2 terms is negligible compared to the costs of computing the MS-mixing term. 

\section{Numerical Examples}

In this section,
we first present the MECIs of ethylene computed by XMS-CASPT2 and compare their structures and energetics with those computed by unc-MRCI.
Second, we investigate the shapes of the potential energy surfaces of a model retinal chromophore (the penta-2,4-dieniminium cation or PSB3) near the MECIs.
Finally, we report the computation of the MECIs for the large molecules
[stilbene and an anionic GFP model chromophore (4-\textit{para}-hydroxybenzylidene-1,2-dimethyl-imidazolin-5-one or \textit{p}HBDI)].
The geometries were optimized using XMS-CASPT2 with the cc-pVDZ basis set and its corresponding JKFIT basis set for density fitting unless mentioned otherwise.
The so-called SS-SR contraction scheme\cite{Vlaisavljevich2016JCTC} with vertical shift (0.2~$E_\mathrm{h}$) was used unless otherwise specified.
We searched for the MECI by the gradient projection method,\cite{Bearpark1994CPL}
that uses the gradient difference and interstate coupling vectors (instead of the full derivative coupling vector).
The flowchart method was used for updating model Hessians for quasi-Newton steps.\cite{Birkholz2016TCA}
For comparison, the MECIs were also optimized using SA-CASSCF.
All of the calculations were performed using the program package \textsc{bagel}.\cite{bagel}

\subsection{Conical Intersections of Ethylene}
First, to assess the accuracy of XMS-CASPT2 against unc-MRCI,\cite{Barbatti2004JCP,Zhang2014JCP} we optimized the MECIs of ethylene.
We adopted the reference SA-CASSCF wave functions from Ref.~\onlinecite{Barbatti2004JCP} with the (2\textit{e},2\textit{o}) active space using the aug-cc-pVDZ basis set.
The three lowest states were averaged in SA-CASSCF. 
The XMS-CASPT2 calculations were performed using a vertical shift of 0.5~$E_\mathrm{h}$.
All of the MECIs reported using unc-MRCI\cite{Barbatti2004JCP,Zhang2014JCP} were also found by XMS-CASPT2: the twisted-pyramid (Py), ethylidene (Et), $C_{3v}$ ethylidene ($C_{3v}$-Et),
H-migration (Hm) ($\mathrm{S}_0$/$\mathrm{S}_1$), and twisted-orthogonal (To) ($\mathrm{S}_1$/$\mathrm{S}_2$) MECIs.
\begin{figure}
\includegraphics[keepaspectratio,width=0.48\textwidth]{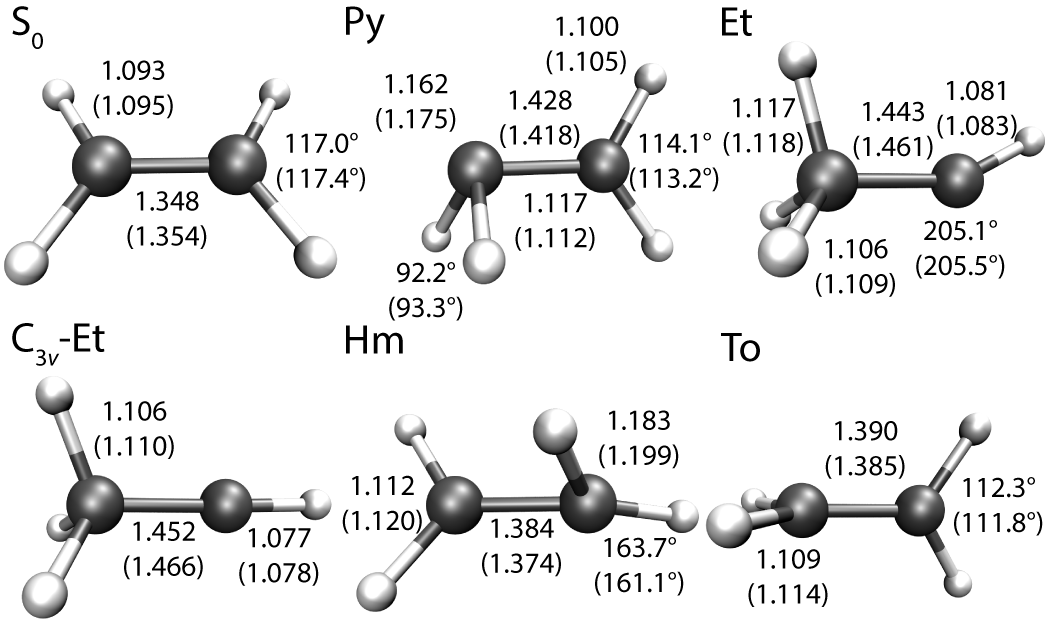}
\caption{
The geometries of the MECIs of ethylene optimized by XMS-CASPT2. The selected geometrical parameters are shown with the unc-MRCI geometrical parameters\cite{Barbatti2004JCP} in the parentheses (bondlengths are shown in \r{A}).\label{ethylene_geometries}
}
\end{figure}
\begin{table}[t]
\caption{Energies of the ground and excited states of ethylene at the $\mathrm{S}_0$ minimum geometry and the MECIs relative to the $\mathrm{S}_0$ minimum energy (eV).
The aug-cc-pVDZ basis set and CAS(2\textit{e},2\textit{o}) are used.\label{ethylene_energies}}
\begin{ruledtabular}
\begin{tabular}{lrrrrrr}
& \multicolumn{3}{c}{XMS-CASPT2}
& \multicolumn{3}{c}{unc-MRCI+Q\textsuperscript{\emph{a}}} \\\hline
& \multicolumn{1}{c}{$\mathrm{S}_0$} & \multicolumn{1}{c}{$\mathrm{S}_1$} & \multicolumn{1}{c}{$\mathrm{S}_2$} 
& \multicolumn{1}{c}{$\mathrm{S}_0$} & \multicolumn{1}{c}{$\mathrm{S}_1$} & \multicolumn{1}{c}{$\mathrm{S}_2$} \\ \hline
$\mathrm{S}_0$ & 0.00  & 7.70  & 13.04  & 0.00  & 7.79  & ...\\
Py  & 4.42  & 4.42  & 9.05    & 4.40  & 4.46  & 9.24 \\
Et  & 4.63  & 4.63  & 5.79    & 4.56  & 4.56  & 5.59 \\
$C_{3v}$-Et& 4.71  & 4.71  & 5.78   & 4.69  & 4.69  & 5.59 \\
Hm  & 5.06  & 5.06  & 8.25    & 5.15  & 5.20  & 8.48 \\
To  & 2.99  & 5.43  & 5.43    & 2.92  & 5.30  & 5.42 \\ 
\end{tabular}
\end{ruledtabular}
\begin{flushleft}\textsuperscript{\emph{a}} Taken from Ref.~\citenum{Barbatti2004JCP}.\end{flushleft}
\end{table}

The geometries of the MECIs and the energies relative
to the $\mathrm{S}_{0}$ minimum are shown and compared with the unc-MRCI results in Fig.~\ref{ethylene_geometries} and in Table~\ref{ethylene_energies}.
They are in good agreement with those computed by unc-MRCI with the Davidson(+Q) correction and the additional restricted active space in the reference function.
This suggests that XMS-CASPT2 can yield results comparable to unc-MRCI+Q in
finding the geometries and energies of MECIs, as in the case for calculating the spectroscopic constants of a range of small molecules.\cite{Azizi2006PCCP}

\subsection{Conical Intersection of PSB3}
Next, we show the potential energy surfaces of cationic PSB3 near the MECIs.
PSB3 is a minimal model for the photoisomerization of protonated Schiff bases,\cite{Gozem2012JCTC,Gozem2014JCTC}
whose conical intersections have been characterized by various electronic structure methods including MS-CASPT2 (using numerical nuclear gradients).\cite{Gozem2012JCTC}
In this work, we optimized the geometries of the MECIs using SA-CASSCF, MS-CASPT2, and XMS-CASPT2.
The two lowest states were averaged in SA-CASSCF.
We used the (6\textit{e},6\textit{o}) active space
consisting of three $\pi$ and three $\pi^{\ast}$ orbitals. The vertical shift of 0.5~$E_\mathrm{h}$ was applied.
The energies at the MECI geometries relative to the \textit{trans} structure in the ground state were found to be 2.36~eV with both MS-CASPT2 and XMS-CASPT2,
which is lower by 0.4~eV than that computed by CASSCF (2.74~eV).
This attests to the importance of dynamical correlation in predicting energies at MECIs.

\begin{figure}[t]
\includegraphics[keepaspectratio,width=0.48\textwidth]{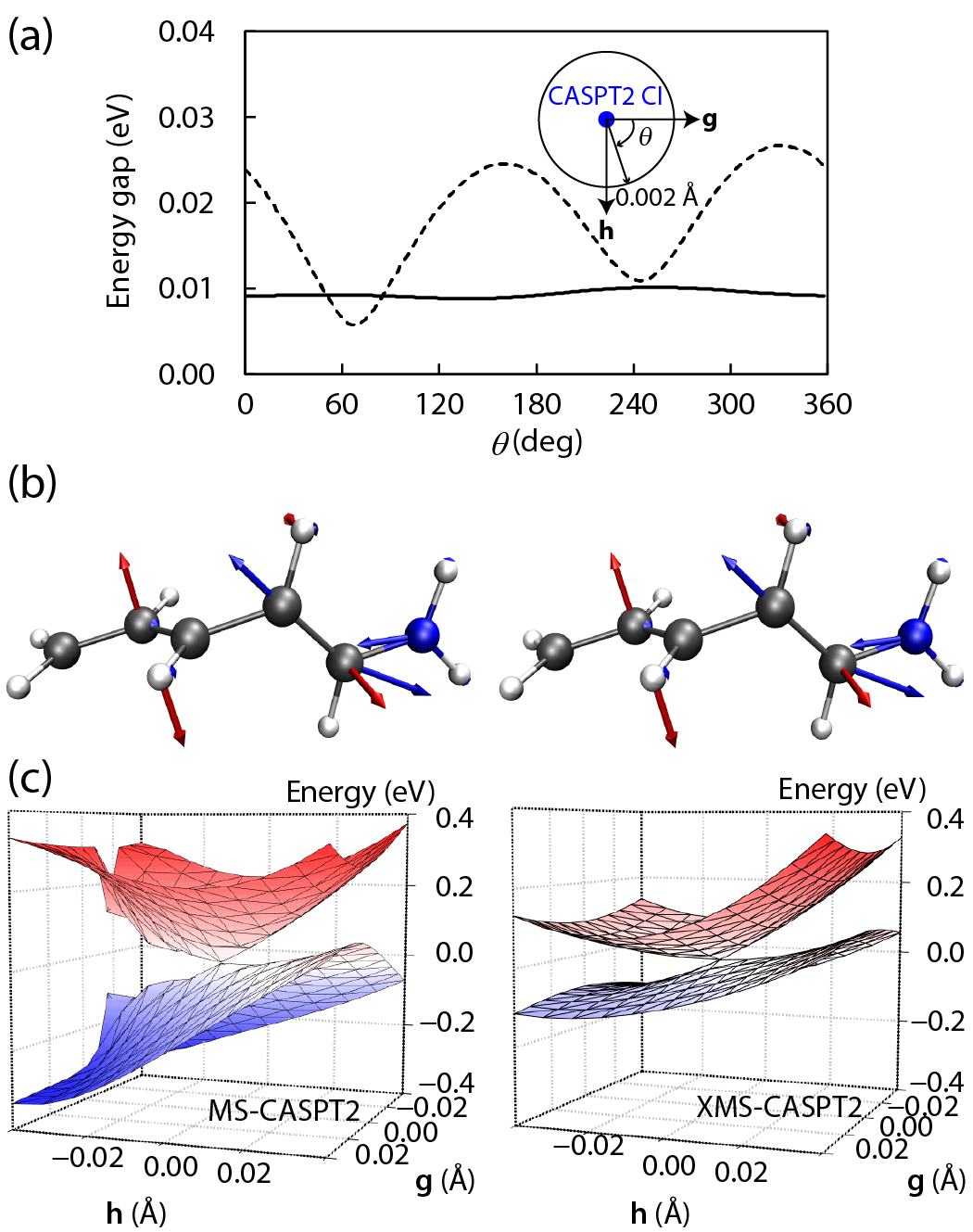}
\caption{
The potential energy surfaces near the MECIs of PSB3.
(a) The $\mathrm{S}_0$ and $\mathrm{S}_1$ energy gaps around the 0.002~\r{A}~radius loop centered at the MECIs, with MS-CASPT2 (dashed) and XMS-CASPT2 (solid). The loop is on the branching plane.
(b) The branching plane at the MECI computed by MS-CASPT2 (left) and XMS-CASPT2 (right). The vectors $\mathbf{g}$ and $\mathbf{h}$ are plotted with blue and red arrows, respectively.
(c) The $\mathrm{S}_0$ and $\mathrm{S}_1$ potential energies on the branching plane, with MS-CASPT2 (left) and XMS-CASPT2 (right). The energy at the MECI is set to zero.\label{psb_ciscan}}
\end{figure}

On the other hand, the shapes of the potential energy surfaces around the MECIs computed by MS-CASPT2 and XMS-CASPT2 were found to differ significantly.
The difference is schematically shown in Fig.~\ref{psb_ciscan}(a) where
the energy gaps are presented along the loop on the branching plane [Fig.~\ref{psb_ciscan}(b)] centered at the MECI,
as in Ref.~\onlinecite{Gozem2014JCTC}.
Note that the branching plane is the plane defined by the gradient difference and interstate coupling vectors ($\mathbf{g}$ and $\mathbf{h}$, respectively).
The radius of the loop was set to 0.002~\AA. We normalized the vectors $\mathbf{g}$ and $\mathbf{h}$ when generating the loop.
The MS-CASPT2 energy gaps showed spurious oscillatory behavior along the loop, as reported in Ref.~\onlinecite{Gozem2014JCTC}, whose amplitude was as large as 0.02~eV.
The gaps computed by XMS-CASPT2 were completely smooth. The amplitude of the oscillation (0.84 kcal/mol) was comparable to that computed by unc-MRCI ($\sim$1~kcal/mol).\cite{Gozem2014JCTC}
The line integral of the XMS-CASPT2 derivative coupling along the same loop\cite{CIbook,Baerbook} was found to be very close to $\pi$ (3.1412).
The potential energy surfaces near the MECIs are also presented in Fig.~\ref{psb_ciscan}(c).
This example stresses the importance of using XMS-CASPT2 (instead of MS-CASPT2) when simulating the electronic structure around conical intersections.

\subsection{Conical Intersections of Stilbene}
Stilbene has been extensively studied computationally as a model compound for photoisomerization around a C=C double bond,\cite{Quenneville2003JPCA,Levine2007ARPC,Minezawa2011JPCA,Ioffe2013JCTC,Lei2014JPCA,Harabuchi2014JPCA}
partly because reliable experimental results are available.\cite{Waldeck1991CR}
The mechanisms for the photodynamics and the locations of the conical intersections have been well characterized.
There are three types of low-lying conical intersections:
one-bond flip (OBF), hula-twist (HT), and 4a,4b-dihydrophenanthrene(DHP)-like conical intersections.
The OBF and HT conical intersections are related to the \textit{cis}--\textit{trans} photoisomerization, and the DHP-like conical intersections are relevant in the formation of DHP, which is a minor product.
We adopted the CASSCF MECI geometries reported in Refs.~\onlinecite{Ioffe2013JCTC} and \onlinecite{Lei2014JPCA} as the starting geometries for the optimizations.
The reference CASSCF wave functions were calculated using two-state averaging with the (6\textit{e},6\textit{o}) active space.

\begin{figure}[t]
\includegraphics[keepaspectratio,width=0.48\textwidth]{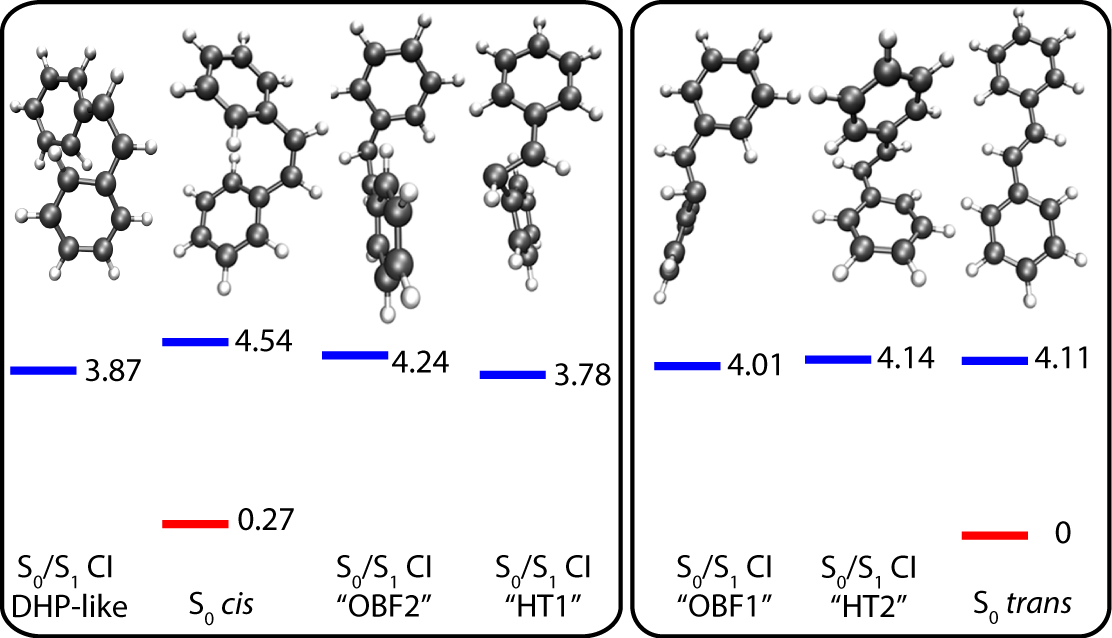}
\caption{
XMS-CASPT2 optimized $\mathrm{S}_0$, $\mathrm{S}_1$ geometries and MECIs of stilbene.
Energies (in eV) are reported relative to the $\mathrm{S}_0$ minimum energy.\label{stilbene_energy}}
\end{figure}

The optimized structures at the $\mathrm{S}_0$ minima and MECIs are shown in Figure \ref{stilbene_energy} along with their energies.
With XMS-CASPT2, two of the MECIs are located below the \textit{trans} Franck--Condon point, and the other two are above that point.
This is in contrast with SA-CASSCF results in which all of the MECIs are about 1~eV below the \textit{trans} Franck--Condon point.
That is, the MECIs are less accessible from the Franck--Condon point when dynamical correlation is considered.
This result is consistent with the previous study based on CASPT2//CASSCF/6-31G.\cite{Lei2014JPCA}

\begin{figure}[t]
\includegraphics[keepaspectratio,width=0.48\textwidth]{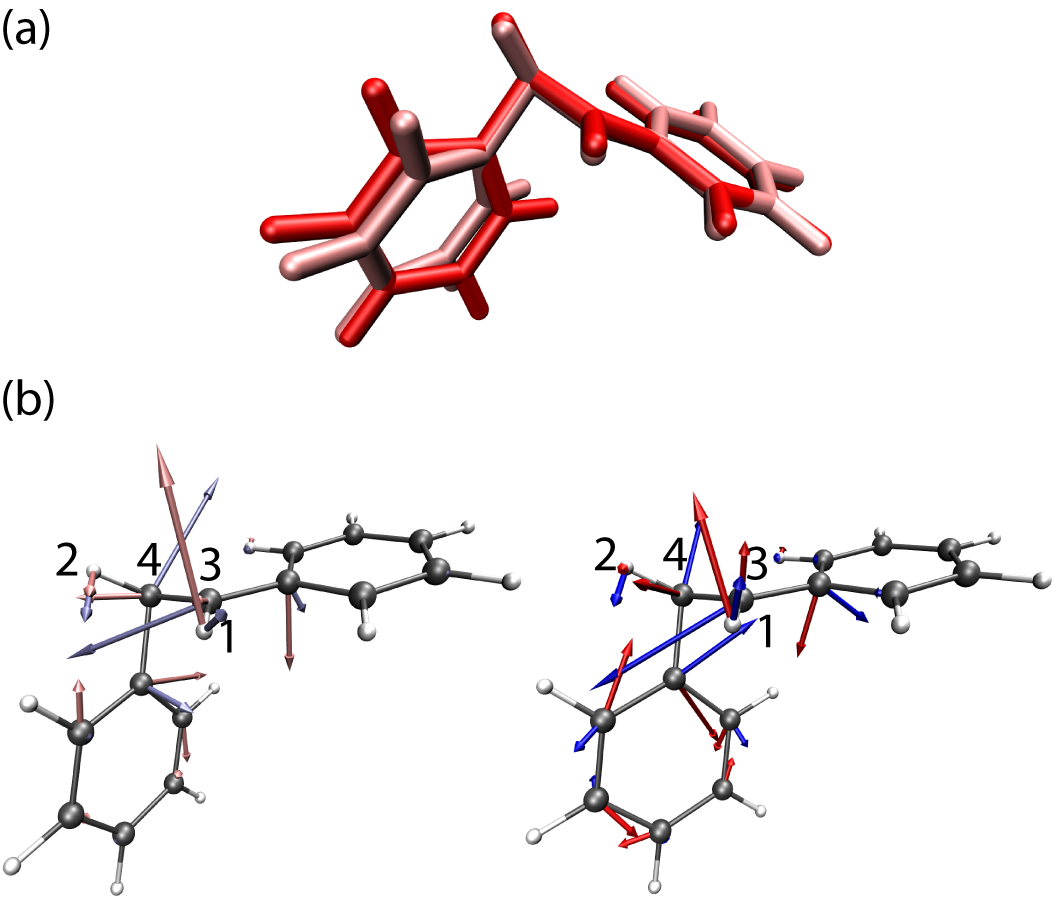}
\caption{
(a) Overlay of the HT1 MECI structures of stilbene optimized by SA-CASSCF (pink) and XMS-CASPT2 (red); (b) The branching plane at the MECI computed by SA-CASSCF (left) and XMS-CASPT2 (right). The vectors $\mathbf{g}$ and $\mathbf{h}$ are plotted with blue and red arrows, respectively.\label{stilbene_overlay}}
\end{figure}

Next, we computed the relative energies of the MECIs using SA-CASSCF and XMS-CASPT2. 
The OBF1 MECIs are 0.21~eV and 0.02~eV lower than the HT1 MECIs with SA-CASSCF using (6\textit{e},6\textit{o}) and (10\textit{e},10\textit{o}) active spaces, which is
different from the XMS-CASPT2 results (0.24~eV higher). This 
is consistent with previous SF-TDDFT\cite{Minezawa2011JPCA} and XMCQDPT2\cite{Ioffe2013JCTC} results.
The DHP-like MECIs are also stabilized by dynamical correlation. These MECIs have higher energies than the other MECIs when dynamical correlation is not included in the calculations;
they are 0.50~eV and 1.18~eV higher than the OBF1 MECIs with SA-CASSCF using (6\textit{e},6\textit{o}) and (10\textit{e},10\textit{o}) active spaces, respectively.
Note that the DHP-like conical intersections become unfavorable when the active space is extended,\cite{Ioffe2013JCTC}
and therefore, further investigations using XMS-CASPT2 and larger active spaces are required to unravel the importance of the DHP-like conical intersections in photodynamics.
The structures of the HT1 MECI of stilbene located using SA-CASSCF and XMS-CASPT2 are shown in Figure \ref{stilbene_overlay} along with the
branching planes. The root mean square deviation of the MECI structures using SA-CASSCF and XMS-CASPT2 was 0.36~\AA~after aligning the two to minimize the deviation,
which is the accuracy that one would achieve when the CASPT2//CASSCF approach is used.

\subsection{Conical Intersections of \textit{p}HBDI}

Next, we show the MECIs of the 4-\textit{para}-hydroxybenzylidene-1,2-dimethyl-imidazolin-5-one (\textit{p}HBDI) anion.
Anionic \textit{p}HBDI is considered to be an emitting species of the 
green fluorescent protein (GFP) and its variants.\cite{Tsien1998ARBC,Weber1999PNAS,Martin2004JACS,Olsen2008JACS,Park2016JACS,vanThor2009CSR,Olsen2010JACS,Polyakov2010JCTC,Acharya2017CR,Morozov2016ACIE,Minezawa2012JCP,Zhang2014ACIE} While GFP exhibits strong fluorescence with a lifetime
on the order of nanoseconds,\cite{Tsien1998ARBC,Martin2004JACS,Olsen2008JACS,vanThor2009CSR,Olsen2010JACS,Polyakov2010JCTC,Park2016JACS,Acharya2017CR} the nonadiabatic transition is known to occur in about a few picoseconds
when the chromophore is not embedded in the protein environment.\cite{Mandal2004JPCB}
As a resonant monomethine dye, it is widely accepted that the anionic GFP chromophore undergoes
nonadiabatic transitions when the chromophore is twisted along the bridge.\cite{Olsen2008JACS,Olsen2010JACS,Olsen2010JCTC,Morozov2016ACIE}
There are two available bridge channels in this molecule, which are the imidazolinone (\textit{I}) and phenolate (\textit{P}) channels,
named after the moiety connected to the bridge bond that twists.\cite{Olsen2010JACS}

We optimized the planar equilibrium geometry of the ground state, the geometries of the \textit{I}- and \textit{P}-twisted minima of the first excited state, and the MECIs between these states near the twisted geometries.
The reference CASSCF wave functions were optimized using three-state averaging with the (4\textit{e},3\textit{o}) active space, which provides a reliable description of twisting of the bridge bond.\cite{Olsen2008JACS,Olsen2009JCP}
\begin{figure}[t]
\includegraphics[keepaspectratio,width=0.48\textwidth]{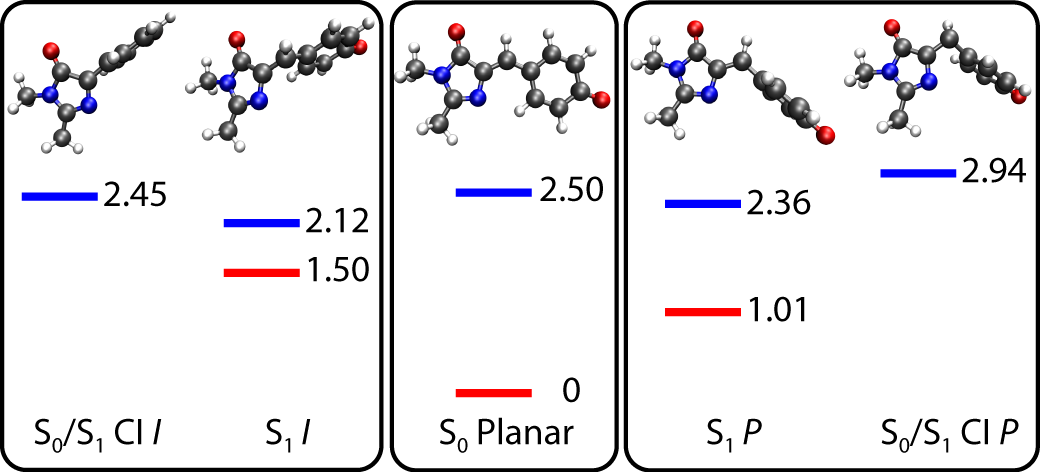}
\caption{XMS-CASPT2 optimized $\mathrm{S}_0$, $\mathrm{S}_1$ geometries and conical intersections of \textit{p}HBDI. Energies (in eV) are reported relative to the $\mathrm{S}_0$ minimum energy.\label{pHBDI_energy}}
\end{figure}
The structures and the energy diagram for these geometries are shown in Figure~\ref{pHBDI_energy}.
The twisted geometries optimized on the first excited state
were found to be lower in energy than the Franck--Condon point by 0.38 and 0.14~eV for the \textit{I}- and \textit{P}-twisted minima, respectively.
However, the conical intersections associated with these twisted geometries are located much higher in energy than the minima;
the \textit{I}-twisted MECI has about the same energy as the Franck--Condon point, and the \textit{P}-twisted MECI lies 0.44 eV above the Franck--Condon point.
This result is in stark contrast to those obtained using SA-CASSCF: the SA-CASSCF energy at the \textit{I}-twisted and \textit{P}-twisted MECIs
are 1.24 and 0.64~eV lower than $\mathrm{S}_1$ energy at the Franck--Condon point.
This suggests that the thermal accessibility of the $\mathrm{S}_0$/$\mathrm{S}_1$ \textit{P}-twisted CI predicted by SA-CASSCF 
is an artifact, because the \textit{P}-twisted MECI lies ca. 10~kcal/mol above the Franck--Condon point on the XMS-CASPT2 surface.
SA-CASSCF results remain qualitatively the same even when the active space is expanded to (12\textit{e},11\textit{o});
the MECIs computed using CAS(12\textit{e},11\textit{o}) were 0.91 and 0.30~eV lower than the Franck--Condon point (see also Ref.~\onlinecite{Polyakov2010JCTC}).
It is notable that this differs qualitatively from the previous CAS(12\textit{e},11\textit{o}) result with two-state averaging,\cite{Martin2004JACS}
which suggests that the choice of the states to be averaged is as important as the choice of the active space.

\begin{figure}[t]
\includegraphics[keepaspectratio,width=0.48\textwidth]{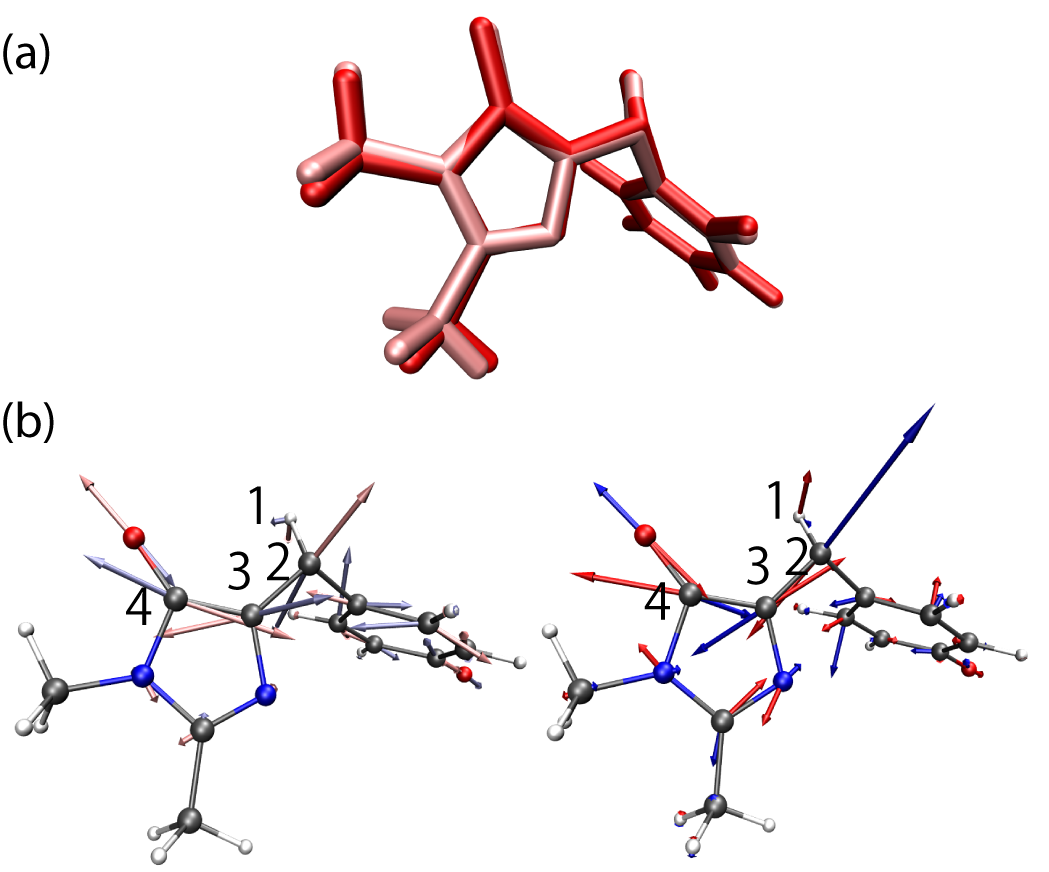}
\caption{
(a) Overlay of the \textit{P}-twisted MECI structures of \textit{p}HBDI optimized by SA-CASSCF (pink) and XMS-CASPT2 (red); (b) The branching plane at the MECI computed by SA-CASSCF (left) and XMS-CASPT2 (right). The vectors $\mathbf{g}$ and $\mathbf{h}$ are plotted with blue and red arrows, respectively.\label{pHBDI_overlay}}
\end{figure}
The structures of the \textit{P}-twisted MECI optimized using SA-CASSCF and XMS-CASPT2 are shown in Figure \ref{pHBDI_overlay}, in which
the full derivative coupling vectors are also shown.
The MECI geometries differ primarily on the bridge: The hydrogen atom on the bridge comes out of the plane of the imidazolinone ring more severely in XMS-CASPT2 than in SA-CASSCF.
The torsional angle around the imidazolinone bridge 
(H1-C2-C3-C4 torsional angle, see Figure~\ref{pHBDI_overlay}) was $-45.1^\circ$ and $-32.0^\circ$ for XMS-CASPT2 and SA-CASSCF, respectively.
Furthermore, the bond length for the imidazolinone bridge (C2-C3 bond) was found to be 1.49 \r{A} (XMS-CASPT2) and 1.45 \r{A} (SA-CASSCF),
indicating that the bridging carbon atom has slightly more $sp^3$ character than $sp^2$ at the \textit{P}-twisted MECI when dynamical correlation is taken into account.
We also note that the hula twist (HT) conical intersection reported in Refs.~\onlinecite{Weber1999PNAS} and \onlinecite{Martin2004JACS} was not found
in our optimization (the optimization converged to the \textit{I}-twisted conical intersection).
Overall, our results imply that including dynamical correlation quantitatively (or, even qualitatively) affects the photochemical dynamics.

\section{Conclusions}

In this work, we have derived the working equations for analytically evaluating the derivative couplings using (X)MS-CASPT2.
The equations have been translated into an efficient computer program as an extension of previously reported analytical gradient programs for (X)MS-CASPT2,
which is included in the {\sc bagel} package\cite{bagel} and publicly available under the GNU General Public License.
The fully internally contracted wave functions were used together with the density fitting approximation.
The computational cost for calculating the derivative couplings was found to be essentially the same as that for computing the nuclear energy gradients for one state.
The energetics at the MECIs were substantially influenced by dynamical correlation.
The optimization of MECIs for ethylene, PSB3, stilbene, and \textit{p}HBDI was presented to demonstrate the versatility of our program.
This finding encourages us to develop a methodology for large-scale direct dynamics in complex systems using a model that takes into account dynamical
correlation contributions such as XMS-CASPT2.
To achieve this goal, efforts to improve our algorithms and implementations (especially for large active spaces) are underway and will be reported in the near future.

\section{Acknowledgments}
The authors thank Dr. Bess Vlaisavljevich for her constructive comments on the manuscript.
The debugging of the SA-CASSCF derivative coupling code in {\sc bagel}, on which this work is based, was facilitated by the existing implementation in {\sc molpro}.\cite{molpro}
This work has been supported by the Air Force Office of Scientific Research Young Investigator Program (Grant No.~FA9550-15-1-0031).
The development of the program infrastructure has been in part supported by National Science Foundation [ACI-1550481 (JWP) and CHE-1351598 (TS)].
T.S. is an Alfred P. Sloan Research Fellow.

\appendix

\section{Numerical XMS-CASPT2 Derivative Coupling}
In part to validate our new analytical derivative coupling program, we have also implemented the code that numerically evaluates (X)MS-CASPT2 derivative coupling. 
In this case, $\mathbf{d}^{\mathrm{XMS},QP}_\mathrm{mix}$ and  $\mathbf{d}^{\mathrm{XMS},QP}_\mathrm{CAS}$
can be combined to the form that resembles SA-CASSCF derivative coupling, using ${\tilde{r}}_{I,Q} = \sum_M \tilde{c}_{I,M} R_{MQ}$, as
\begin{subequations}
\begin{align}
\mathbf{d}^{\mathrm{XMS},QP}_\mathrm{mix} + \mathbf{d}^{\mathrm{XMS},QP}_\mathrm{CAS}
=\sum_{I} {\tilde{r}}_{I,Q} \frac{d {\tilde{r}}_{I,P}}{d \mathbf{X}} + 
\sum_{IJ} {\tilde{r}}_{I,Q} {\tilde{r}}_{J,P} \braket{I}{\frac{d J}{d \mathbf{X}}},
\end{align}
\end{subequations}
and is evaluated numerically as\cite{Galloy1977JCP}
\begin{align}
&\left(\mathbf{d}^{\mathrm{XMS},QP}_\mathrm{mix} + \mathbf{d}^{\mathrm{XMS},QP}_\mathrm{CAS} \right)_{X}\nonumber\\
&\quad =
\sum_{I} {\tilde{r}}_{I,Q} \frac{\Delta{\tilde{r}}_{I,P} }{\Delta X}
+ \sum_{rs} {\gamma}^{QP}_{rs} \left[ \sum_{\mu\nu}C_{\mu r} S_{\mu\nu} \frac{\Delta C_{\nu s}}{\Delta X} + \sigma_{rs}^{X}\right].
\end{align}
The PT2 term is calculated using the following expression:
\begin{align}
\left( \mathbf{d}^{\mathrm{XMS},QP}_\mathrm{PT2} \right)_X
&= \sum_{KL} \mathcal{R}^{QP}_{KL} \sum_{rs} \langle \Phi^{(1)}_K|\hat{E}_{rs}|L\rangle
\nonumber\\ &\times
\left[  \sum_{\mu\nu}C_{\mu r} S_{\mu\nu} 
\frac{\Delta C_{\nu s}}{\Delta X}
+ \sigma_{rs}^{X}  \right].
\end{align}
The derivatives ($\Delta {\tilde{r}}_{I,P} / \Delta X$ and  $\Delta C_{\mu r}/\Delta X$)
were calculated by means of the finite difference formula.
The numerical derivative couplings agreed with those analytically evaluated.


\begin{thebibliography}{78}%
\makeatletter
\providecommand \@ifxundefined [1]{%
 \@ifx{#1\undefined}
}%
\providecommand \@ifnum [1]{%
 \ifnum #1\expandafter \@firstoftwo
 \else \expandafter \@secondoftwo
 \fi
}%
\providecommand \@ifx [1]{%
 \ifx #1\expandafter \@firstoftwo
 \else \expandafter \@secondoftwo
 \fi
}%
\providecommand \natexlab [1]{#1}%
\providecommand \enquote  [1]{``#1''}%
\providecommand \bibnamefont  [1]{#1}%
\providecommand \bibfnamefont [1]{#1}%
\providecommand \citenamefont [1]{#1}%
\providecommand \href@noop [0]{\@secondoftwo}%
\providecommand \href [0]{\begingroup \@sanitize@url \@href}%
\providecommand \@href[1]{\@@startlink{#1}\@@href}%
\providecommand \@@href[1]{\endgroup#1\@@endlink}%
\providecommand \@sanitize@url [0]{\catcode `\\12\catcode `\$12\catcode
  `\&12\catcode `\#12\catcode `\^12\catcode `\_12\catcode `\%12\relax}%
\providecommand \@@startlink[1]{}%
\providecommand \@@endlink[0]{}%
\providecommand \url  [0]{\begingroup\@sanitize@url \@url }%
\providecommand \@url [1]{\endgroup\@href {#1}{\urlprefix }}%
\providecommand \urlprefix  [0]{URL }%
\providecommand \Eprint [0]{\href }%
\providecommand \doibase [0]{http://dx.doi.org/}%
\providecommand \selectlanguage [0]{\@gobble}%
\providecommand \bibinfo  [0]{\@secondoftwo}%
\providecommand \bibfield  [0]{\@secondoftwo}%
\providecommand \translation [1]{[#1]}%
\providecommand \BibitemOpen [0]{}%
\providecommand \bibitemStop [0]{}%
\providecommand \bibitemNoStop [0]{.\EOS\space}%
\providecommand \EOS [0]{\spacefactor3000\relax}%
\providecommand \BibitemShut  [1]{\csname bibitem#1\endcsname}%
\let\auto@bib@innerbib\@empty
\bibitem [{\citenamefont {Valeur}(2002)}]{Valeurbook}%
  \BibitemOpen
  \bibfield  {author} {\bibinfo {author} {\bibfnamefont {B.}~\bibnamefont
  {Valeur}},\ }\href@noop {} {\emph {\bibinfo {title} {{M}olecular
  {F}luorescence: {P}rinciples and {A}pplications}}}\ (\bibinfo  {publisher}
  {Wiley-VCH},\ \bibinfo {address} {Weinheim, Germany},\ \bibinfo {year}
  {2002})\ pp.\ \bibinfo {pages} {20--124}\BibitemShut {NoStop}%
\bibitem [{\citenamefont {Levine}\ and\ \citenamefont
  {{Mart\'inez}}(2007)}]{Levine2007ARPC}%
  \BibitemOpen
  \bibfield  {author} {\bibinfo {author} {\bibfnamefont {B.~G.}\ \bibnamefont
  {Levine}}\ and\ \bibinfo {author} {\bibfnamefont {T.~J.}\ \bibnamefont
  {{Mart\'inez}}},\ }\href@noop {} {\bibfield  {journal} {\bibinfo  {journal}
  {Annu. Rev. Phys. Chem.}\ }\textbf {\bibinfo {volume} {58}},\ \bibinfo
  {pages} {613} (\bibinfo {year} {2007})}\BibitemShut {NoStop}%
\bibitem [{\citenamefont {Erbas-Cakmak}\ \emph {et~al.}(2015)\citenamefont
  {Erbas-Cakmak}, \citenamefont {Leigh}, \citenamefont {McTernan},\ and\
  \citenamefont {Nussbaumer}}]{Erbak-Cakmak2015CR}%
  \BibitemOpen
  \bibfield  {author} {\bibinfo {author} {\bibfnamefont {S.}~\bibnamefont
  {Erbas-Cakmak}}, \bibinfo {author} {\bibfnamefont {D.~A.}\ \bibnamefont
  {Leigh}}, \bibinfo {author} {\bibfnamefont {C.~T.}\ \bibnamefont {McTernan}},
  \ and\ \bibinfo {author} {\bibfnamefont {A.~L.}\ \bibnamefont {Nussbaumer}},\
  }\href@noop {} {\bibfield  {journal} {\bibinfo  {journal} {Chem. Rev.}\
  }\textbf {\bibinfo {volume} {115}},\ \bibinfo {pages} {10081} (\bibinfo
  {year} {2015})}\BibitemShut {NoStop}%
\bibitem [{\citenamefont {Matsika}\ and\ \citenamefont
  {Krause}(2011)}]{Matsika2011ARPC}%
  \BibitemOpen
  \bibfield  {author} {\bibinfo {author} {\bibfnamefont {S.}~\bibnamefont
  {Matsika}}\ and\ \bibinfo {author} {\bibfnamefont {P.}~\bibnamefont
  {Krause}},\ }\href@noop {} {\bibfield  {journal} {\bibinfo  {journal} {Annu.
  Rev. Phys. Chem}\ }\textbf {\bibinfo {volume} {62}},\ \bibinfo {pages} {621}
  (\bibinfo {year} {2011})}\BibitemShut {NoStop}%
\bibitem [{\citenamefont {Matsika}\ \emph {et~al.}(2014)\citenamefont
  {Matsika}, \citenamefont {Feng}, \citenamefont {Luzanov},\ and\ \citenamefont
  {Krylov}}]{Matsika2014JPCA}%
  \BibitemOpen
  \bibfield  {author} {\bibinfo {author} {\bibfnamefont {S.}~\bibnamefont
  {Matsika}}, \bibinfo {author} {\bibfnamefont {X.}~\bibnamefont {Feng}},
  \bibinfo {author} {\bibfnamefont {A.~V.}\ \bibnamefont {Luzanov}}, \ and\
  \bibinfo {author} {\bibfnamefont {A.~I.}\ \bibnamefont {Krylov}},\
  }\href@noop {} {\bibfield  {journal} {\bibinfo  {journal} {J. Phys. Chem. A}\
  }\textbf {\bibinfo {volume} {118}},\ \bibinfo {pages} {11943} (\bibinfo
  {year} {2014})}\BibitemShut {NoStop}%
\bibitem [{\citenamefont {Olsen}\ \emph {et~al.}(2010)\citenamefont {Olsen},
  \citenamefont {Lamothe},\ and\ \citenamefont
  {{Mart\'{i}nez}}}]{Olsen2010JACS}%
  \BibitemOpen
  \bibfield  {author} {\bibinfo {author} {\bibfnamefont {S.}~\bibnamefont
  {Olsen}}, \bibinfo {author} {\bibfnamefont {K.}~\bibnamefont {Lamothe}}, \
  and\ \bibinfo {author} {\bibfnamefont {T.~J.}\ \bibnamefont
  {{Mart\'{i}nez}}},\ }\href@noop {} {\bibfield  {journal} {\bibinfo  {journal}
  {J. Am. Chem. Soc.}\ }\textbf {\bibinfo {volume} {132}},\ \bibinfo {pages}
  {1192 } (\bibinfo {year} {2010})}\BibitemShut {NoStop}%
\bibitem [{\citenamefont {Lengsfield}\ \emph {et~al.}(1984)\citenamefont
  {Lengsfield}, \citenamefont {Saxe},\ and\ \citenamefont
  {Yarkony}}]{Lengsfield1984JCP}%
  \BibitemOpen
  \bibfield  {author} {\bibinfo {author} {\bibfnamefont {B.~H.}\ \bibnamefont
  {Lengsfield}, \bibfnamefont {III}}, \bibinfo {author} {\bibfnamefont
  {P.}~\bibnamefont {Saxe}}, \ and\ \bibinfo {author} {\bibfnamefont {D.~R.}\
  \bibnamefont {Yarkony}},\ }\href@noop {} {\bibfield  {journal} {\bibinfo
  {journal} {J. Chem. Phys.}\ }\textbf {\bibinfo {volume} {81}},\ \bibinfo
  {pages} {4549} (\bibinfo {year} {1984})}\BibitemShut {NoStop}%
\bibitem [{\citenamefont {Lengsfield}\ and\ \citenamefont
  {Yarkony}(1992)}]{Lengsfield1992ACP}%
  \BibitemOpen
  \bibfield  {author} {\bibinfo {author} {\bibfnamefont {B.~H.}\ \bibnamefont
  {Lengsfield}, \bibfnamefont {III}}\ and\ \bibinfo {author} {\bibfnamefont
  {D.~R.}\ \bibnamefont {Yarkony}},\ }\href@noop {} {\bibfield  {journal}
  {\bibinfo  {journal} {Adv. Chem. Phys.}\ }\textbf {\bibinfo {volume} {82}},\
  \bibinfo {pages} {1} (\bibinfo {year} {1992})}\BibitemShut {NoStop}%
\bibitem [{\citenamefont {Galvan}\ \emph {et~al.}(2016)\citenamefont {Galvan},
  \citenamefont {Delcey}, \citenamefont {Pedersen}, \citenamefont {Aquilante},\
  and\ \citenamefont {Lindh}}]{Galvan2016JCTC}%
  \BibitemOpen
  \bibfield  {author} {\bibinfo {author} {\bibfnamefont {I.~F.}\ \bibnamefont
  {Galvan}}, \bibinfo {author} {\bibfnamefont {M.~G.}\ \bibnamefont {Delcey}},
  \bibinfo {author} {\bibfnamefont {T.~B.}\ \bibnamefont {Pedersen}}, \bibinfo
  {author} {\bibfnamefont {F.}~\bibnamefont {Aquilante}}, \ and\ \bibinfo
  {author} {\bibfnamefont {R.}~\bibnamefont {Lindh}},\ }\href@noop {}
  {\bibfield  {journal} {\bibinfo  {journal} {J. Chem. Theory Comput.}\
  }\textbf {\bibinfo {volume} {12}},\ \bibinfo {pages} {3636} (\bibinfo {year}
  {2016})}\BibitemShut {NoStop}%
\bibitem [{\citenamefont {Lischka}\ \emph {et~al.}(2004)\citenamefont
  {Lischka}, \citenamefont {Dallos}, \citenamefont {Szalay}, \citenamefont
  {Yarkony},\ and\ \citenamefont {Shepard}}]{Lischka2004JCP}%
  \BibitemOpen
  \bibfield  {author} {\bibinfo {author} {\bibfnamefont {H.}~\bibnamefont
  {Lischka}}, \bibinfo {author} {\bibfnamefont {M.}~\bibnamefont {Dallos}},
  \bibinfo {author} {\bibfnamefont {P.~G.}\ \bibnamefont {Szalay}}, \bibinfo
  {author} {\bibfnamefont {D.~R.}\ \bibnamefont {Yarkony}}, \ and\ \bibinfo
  {author} {\bibfnamefont {R.}~\bibnamefont {Shepard}},\ }\href@noop {}
  {\bibfield  {journal} {\bibinfo  {journal} {J. Chem. Phys.}\ }\textbf
  {\bibinfo {volume} {120}},\ \bibinfo {pages} {7322} (\bibinfo {year}
  {2004})}\BibitemShut {NoStop}%
\bibitem [{\citenamefont {Barbatti}\ and\ \citenamefont
  {Lischka}(2008)}]{Barbatti2008JACS}%
  \BibitemOpen
  \bibfield  {author} {\bibinfo {author} {\bibfnamefont {M.}~\bibnamefont
  {Barbatti}}\ and\ \bibinfo {author} {\bibfnamefont {H.}~\bibnamefont
  {Lischka}},\ }\href@noop {} {\bibfield  {journal} {\bibinfo  {journal} {J.
  Am. Chem. Soc.}\ }\textbf {\bibinfo {volume} {130}},\ \bibinfo {pages} {6831}
  (\bibinfo {year} {2008})}\BibitemShut {NoStop}%
\bibitem [{\citenamefont {Tajti}\ and\ \citenamefont
  {Szalay}(2009)}]{Tajti2009JCP}%
  \BibitemOpen
  \bibfield  {author} {\bibinfo {author} {\bibfnamefont {A.}~\bibnamefont
  {Tajti}}\ and\ \bibinfo {author} {\bibfnamefont {P.~G.}\ \bibnamefont
  {Szalay}},\ }\href@noop {} {\bibfield  {journal} {\bibinfo  {journal} {J.
  Chem. Phys.}\ }\textbf {\bibinfo {volume} {131}},\ \bibinfo {pages} {124104}
  (\bibinfo {year} {2009})}\BibitemShut {NoStop}%
\bibitem [{\citenamefont {Christiansen}(1999)}]{Christiansen1999JCP2}%
  \BibitemOpen
  \bibfield  {author} {\bibinfo {author} {\bibfnamefont {O.}~\bibnamefont
  {Christiansen}},\ }\href@noop {} {\bibfield  {journal} {\bibinfo  {journal}
  {J. Chem. Phys.}\ }\textbf {\bibinfo {volume} {110}},\ \bibinfo {pages} {711}
  (\bibinfo {year} {1999})}\BibitemShut {NoStop}%
\bibitem [{\citenamefont {Ichino}\ \emph {et~al.}(2009)\citenamefont {Ichino},
  \citenamefont {Gauss},\ and\ \citenamefont {Stanton}}]{Ichino2009JCP}%
  \BibitemOpen
  \bibfield  {author} {\bibinfo {author} {\bibfnamefont {T.}~\bibnamefont
  {Ichino}}, \bibinfo {author} {\bibfnamefont {J.}~\bibnamefont {Gauss}}, \
  and\ \bibinfo {author} {\bibfnamefont {J.~F.}\ \bibnamefont {Stanton}},\
  }\href@noop {} {\bibfield  {journal} {\bibinfo  {journal} {J. Chem. Phys.}\
  }\textbf {\bibinfo {volume} {130}},\ \bibinfo {pages} {174105} (\bibinfo
  {year} {2009})}\BibitemShut {NoStop}%
\bibitem [{\citenamefont {Fatehi}\ \emph {et~al.}(2011)\citenamefont {Fatehi},
  \citenamefont {Alguire}, \citenamefont {Shao},\ and\ \citenamefont
  {Subotnik}}]{Fatehi2011JCP}%
  \BibitemOpen
  \bibfield  {author} {\bibinfo {author} {\bibfnamefont {S.}~\bibnamefont
  {Fatehi}}, \bibinfo {author} {\bibfnamefont {E.}~\bibnamefont {Alguire}},
  \bibinfo {author} {\bibfnamefont {Y.}~\bibnamefont {Shao}}, \ and\ \bibinfo
  {author} {\bibfnamefont {J.~E.}\ \bibnamefont {Subotnik}},\ }\href@noop {}
  {\bibfield  {journal} {\bibinfo  {journal} {J. Chem. Phys.}\ }\textbf
  {\bibinfo {volume} {135}},\ \bibinfo {pages} {234105} (\bibinfo {year}
  {2011})}\BibitemShut {NoStop}%
\bibitem [{\citenamefont {Chernyak}\ and\ \citenamefont
  {Mukamel}(2000)}]{Chernyak2000JCP}%
  \BibitemOpen
  \bibfield  {author} {\bibinfo {author} {\bibfnamefont {V.}~\bibnamefont
  {Chernyak}}\ and\ \bibinfo {author} {\bibfnamefont {S.}~\bibnamefont
  {Mukamel}},\ }\href@noop {} {\bibfield  {journal} {\bibinfo  {journal} {J.
  Chem. Phys.}\ }\textbf {\bibinfo {volume} {112}},\ \bibinfo {pages} {3572}
  (\bibinfo {year} {2000})}\BibitemShut {NoStop}%
\bibitem [{\citenamefont {Send}\ and\ \citenamefont
  {Furche}(2010)}]{Send2010JCP}%
  \BibitemOpen
  \bibfield  {author} {\bibinfo {author} {\bibfnamefont {R.}~\bibnamefont
  {Send}}\ and\ \bibinfo {author} {\bibfnamefont {F.}~\bibnamefont {Furche}},\
  }\href@noop {} {\bibfield  {journal} {\bibinfo  {journal} {J. Chem. Phys.}\
  }\textbf {\bibinfo {volume} {132}},\ \bibinfo {pages} {044107} (\bibinfo
  {year} {2010})}\BibitemShut {NoStop}%
\bibitem [{\citenamefont {Zhang}\ and\ \citenamefont
  {Herbert}(2014)}]{Zhang2014JCP}%
  \BibitemOpen
  \bibfield  {author} {\bibinfo {author} {\bibfnamefont {X.}~\bibnamefont
  {Zhang}}\ and\ \bibinfo {author} {\bibfnamefont {J.~M.}\ \bibnamefont
  {Herbert}},\ }\href@noop {} {\bibfield  {journal} {\bibinfo  {journal} {J.
  Chem. Phys.}\ }\textbf {\bibinfo {volume} {141}},\ \bibinfo {pages} {064104}
  (\bibinfo {year} {2014})}\BibitemShut {NoStop}%
\bibitem [{\citenamefont {Zhang}\ and\ \citenamefont
  {Herbert}(2015)}]{Zhang2015JCP}%
  \BibitemOpen
  \bibfield  {author} {\bibinfo {author} {\bibfnamefont {X.}~\bibnamefont
  {Zhang}}\ and\ \bibinfo {author} {\bibfnamefont {J.~M.}\ \bibnamefont
  {Herbert}},\ }\href@noop {} {\bibfield  {journal} {\bibinfo  {journal} {J.
  Chem. Phys.}\ }\textbf {\bibinfo {volume} {143}},\ \bibinfo {pages} {234107}
  (\bibinfo {year} {2015})}\BibitemShut {NoStop}%
\bibitem [{\citenamefont {Herbert}\ \emph {et~al.}(2016)\citenamefont
  {Herbert}, \citenamefont {Zhang}, \citenamefont {Morrison},\ and\
  \citenamefont {Liu}}]{Herbert2016ACR}%
  \BibitemOpen
  \bibfield  {author} {\bibinfo {author} {\bibfnamefont {J.~M.}\ \bibnamefont
  {Herbert}}, \bibinfo {author} {\bibfnamefont {X.}~\bibnamefont {Zhang}},
  \bibinfo {author} {\bibfnamefont {A.~F.}\ \bibnamefont {Morrison}}, \ and\
  \bibinfo {author} {\bibfnamefont {J.}~\bibnamefont {Liu}},\ }\href@noop {}
  {\bibfield  {journal} {\bibinfo  {journal} {Acc. Chem. Res.}\ }\textbf
  {\bibinfo {volume} {49}},\ \bibinfo {pages} {931} (\bibinfo {year}
  {2016})}\BibitemShut {NoStop}%
\bibitem [{Note1()}]{Note1}%
  \BibitemOpen
  \bibinfo {note} {The terms 'derivative coupling' and 'nonadiabatic coupling'
  are conventionally used interchangeably, except for some recent reports, such
  as Ref.~\protect \citenum {Galvan2016JCTC}}\BibitemShut {NoStop}%
\bibitem [{\citenamefont {Tully}(1990)}]{Tully1990JCP}%
  \BibitemOpen
  \bibfield  {author} {\bibinfo {author} {\bibfnamefont {J.~C.}\ \bibnamefont
  {Tully}},\ }\href@noop {} {\bibfield  {journal} {\bibinfo  {journal} {J.
  Chem. Phys.}\ }\textbf {\bibinfo {volume} {93}},\ \bibinfo {pages} {1061}
  (\bibinfo {year} {1990})}\BibitemShut {NoStop}%
\bibitem [{\citenamefont {Nelson}\ \emph {et~al.}(2014)\citenamefont {Nelson},
  \citenamefont {Fernandez-Alberti}, \citenamefont {Roitberg},\ and\
  \citenamefont {Tretiak}}]{Nelson2014ACR}%
  \BibitemOpen
  \bibfield  {author} {\bibinfo {author} {\bibfnamefont {T.}~\bibnamefont
  {Nelson}}, \bibinfo {author} {\bibfnamefont {S.}~\bibnamefont
  {Fernandez-Alberti}}, \bibinfo {author} {\bibfnamefont {A.~E.}\ \bibnamefont
  {Roitberg}}, \ and\ \bibinfo {author} {\bibfnamefont {S.}~\bibnamefont
  {Tretiak}},\ }\href@noop {} {\bibfield  {journal} {\bibinfo  {journal} {Acc.
  Chem. Res.}\ }\textbf {\bibinfo {volume} {47}},\ \bibinfo {pages} {1155}
  (\bibinfo {year} {2014})}\BibitemShut {NoStop}%
\bibitem [{\citenamefont {Tavernelli}(2015)}]{Tavernelli2015ACR}%
  \BibitemOpen
  \bibfield  {author} {\bibinfo {author} {\bibfnamefont {I.}~\bibnamefont
  {Tavernelli}},\ }\href@noop {} {\bibfield  {journal} {\bibinfo  {journal}
  {Acc. Chem. Res.}\ }\textbf {\bibinfo {volume} {48}},\ \bibinfo {pages} {792}
  (\bibinfo {year} {2015})}\BibitemShut {NoStop}%
\bibitem [{\citenamefont {Domcke}\ \emph {et~al.}(2004)\citenamefont {Domcke},
  \citenamefont {Yarkony},\ and\ \citenamefont {K{\"o}ppel}}]{CIbook}%
  \BibitemOpen
  \bibinfo {editor} {\bibfnamefont {W.}~\bibnamefont {Domcke}}, \bibinfo
  {editor} {\bibfnamefont {D.~R.}\ \bibnamefont {Yarkony}}, \ and\ \bibinfo
  {editor} {\bibfnamefont {H.}~\bibnamefont {K{\"o}ppel}},\ eds.,\ \href@noop
  {} {\emph {\bibinfo {title} {{C}onical {I}ntersections: {T}heory,
  {C}omputation and {E}xperiment}}}\ (\bibinfo  {publisher} {World
  Scientific},\ \bibinfo {address} {Singapore},\ \bibinfo {year} {2004})\ pp.\
  \bibinfo {pages} {3--174}\BibitemShut {NoStop}%
\bibitem [{\citenamefont {Baer}(2006)}]{Baerbook}%
  \BibitemOpen
  \bibfield  {author} {\bibinfo {author} {\bibfnamefont {M.}~\bibnamefont
  {Baer}},\ }\href@noop {} {\emph {\bibinfo {title} {{B}eyond
  {B}orn--{O}ppenheimer: {E}lectronic {N}onadiabatic {C}oupling {T}erms and
  {C}onical {I}ntersections}}}\ (\bibinfo  {publisher} {John Wiley \& Sons},\
  \bibinfo {address} {Hoboken, NJ},\ \bibinfo {year} {2006})\ pp.\ \bibinfo
  {pages} {26--138}\BibitemShut {NoStop}%
\bibitem [{\citenamefont {Bearpark}\ \emph {et~al.}(1994)\citenamefont
  {Bearpark}, \citenamefont {Robb},\ and\ \citenamefont
  {Schlegel}}]{Bearpark1994CPL}%
  \BibitemOpen
  \bibfield  {author} {\bibinfo {author} {\bibfnamefont {M.~J.}\ \bibnamefont
  {Bearpark}}, \bibinfo {author} {\bibfnamefont {M.~A.}\ \bibnamefont {Robb}},
  \ and\ \bibinfo {author} {\bibfnamefont {H.~B.}\ \bibnamefont {Schlegel}},\
  }\href@noop {} {\bibfield  {journal} {\bibinfo  {journal} {Chem. Phys.
  Lett.}\ }\textbf {\bibinfo {volume} {223}},\ \bibinfo {pages} {269} (\bibinfo
  {year} {1994})}\BibitemShut {NoStop}%
\bibitem [{\citenamefont {Manaa}\ and\ \citenamefont
  {Yarkony}(1993)}]{Manaa1993JCP}%
  \BibitemOpen
  \bibfield  {author} {\bibinfo {author} {\bibfnamefont {M.~R.}\ \bibnamefont
  {Manaa}}\ and\ \bibinfo {author} {\bibfnamefont {D.~R.}\ \bibnamefont
  {Yarkony}},\ }\href@noop {} {\bibfield  {journal} {\bibinfo  {journal} {J.
  Chem. Phys.}\ }\textbf {\bibinfo {volume} {99}},\ \bibinfo {pages} {5251}
  (\bibinfo {year} {1993})}\BibitemShut {NoStop}%
\bibitem [{\citenamefont {Yarkony}(1996)}]{Yarkony1996RMP}%
  \BibitemOpen
  \bibfield  {author} {\bibinfo {author} {\bibfnamefont {D.~R.}\ \bibnamefont
  {Yarkony}},\ }\href@noop {} {\bibfield  {journal} {\bibinfo  {journal} {{Rev.
  Mod. Phys.}}\ }\textbf {\bibinfo {volume} {68}},\ \bibinfo {pages} {985}
  (\bibinfo {year} {1996})}\BibitemShut {NoStop}%
\bibitem [{\citenamefont {Levine}\ \emph {et~al.}(2006)\citenamefont {Levine},
  \citenamefont {Ko}, \citenamefont {Quenneville},\ and\ \citenamefont
  {{Mart\'{i}nez}}}]{Levine2006MP}%
  \BibitemOpen
  \bibfield  {author} {\bibinfo {author} {\bibfnamefont {B.~G.}\ \bibnamefont
  {Levine}}, \bibinfo {author} {\bibfnamefont {C.}~\bibnamefont {Ko}}, \bibinfo
  {author} {\bibfnamefont {J.}~\bibnamefont {Quenneville}}, \ and\ \bibinfo
  {author} {\bibfnamefont {T.~J.}\ \bibnamefont {{Mart\'{i}nez}}},\ }\href@noop
  {} {\bibfield  {journal} {\bibinfo  {journal} {Mol. Phys.}\ }\textbf
  {\bibinfo {volume} {104}},\ \bibinfo {pages} {1039} (\bibinfo {year}
  {2006})}\BibitemShut {NoStop}%
\bibitem [{\citenamefont {Huix-Rotllant}\ \emph {et~al.}(2013)\citenamefont
  {Huix-Rotllant}, \citenamefont {Filatov}, \citenamefont {Gozem},
  \citenamefont {Schapiro}, \citenamefont {Olivucci},\ and\ \citenamefont
  {{Ferr\'{e}}}}]{Huix-Rotllant2013JCTC}%
  \BibitemOpen
  \bibfield  {author} {\bibinfo {author} {\bibfnamefont {M.}~\bibnamefont
  {Huix-Rotllant}}, \bibinfo {author} {\bibfnamefont {M.}~\bibnamefont
  {Filatov}}, \bibinfo {author} {\bibfnamefont {S.}~\bibnamefont {Gozem}},
  \bibinfo {author} {\bibfnamefont {I.}~\bibnamefont {Schapiro}}, \bibinfo
  {author} {\bibfnamefont {M.}~\bibnamefont {Olivucci}}, \ and\ \bibinfo
  {author} {\bibfnamefont {N.}~\bibnamefont {{Ferr\'{e}}}},\ }\href@noop {}
  {\bibfield  {journal} {\bibinfo  {journal} {J. Chem. Theory Comput.}\
  }\textbf {\bibinfo {volume} {9}},\ \bibinfo {pages} {3917} (\bibinfo {year}
  {2013})}\BibitemShut {NoStop}%
\bibitem [{\citenamefont {Li}\ \emph {et~al.}(2014)\citenamefont {Li},
  \citenamefont {Marenich}, \citenamefont {Xu},\ and\ \citenamefont
  {Truhlar}}]{Li2014JPCL}%
  \BibitemOpen
  \bibfield  {author} {\bibinfo {author} {\bibfnamefont {S.~L.}\ \bibnamefont
  {Li}}, \bibinfo {author} {\bibfnamefont {A.~V.}\ \bibnamefont {Marenich}},
  \bibinfo {author} {\bibfnamefont {X.}~\bibnamefont {Xu}}, \ and\ \bibinfo
  {author} {\bibfnamefont {D.~G.}\ \bibnamefont {Truhlar}},\ }\href@noop {}
  {\bibfield  {journal} {\bibinfo  {journal} {J. Phys. Chem. Lett.}\ }\textbf
  {\bibinfo {volume} {5}},\ \bibinfo {pages} {322} (\bibinfo {year}
  {2014})}\BibitemShut {NoStop}%
\bibitem [{\citenamefont {Tuna}\ \emph {et~al.}(2015)\citenamefont {Tuna},
  \citenamefont {Lefrancois}, \citenamefont {Wola\'{n}ski}, \citenamefont
  {Gozem}, \citenamefont {Schapiro}, \citenamefont {{Andruni\'{o}w}},
  \citenamefont {Dreuw},\ and\ \citenamefont {Olivucci}}]{Tuna2015JCTC}%
  \BibitemOpen
  \bibfield  {author} {\bibinfo {author} {\bibfnamefont {D.}~\bibnamefont
  {Tuna}}, \bibinfo {author} {\bibfnamefont {D.}~\bibnamefont {Lefrancois}},
  \bibinfo {author} {\bibfnamefont {{\L}.}~\bibnamefont {Wola\'{n}ski}},
  \bibinfo {author} {\bibfnamefont {S.}~\bibnamefont {Gozem}}, \bibinfo
  {author} {\bibfnamefont {I.}~\bibnamefont {Schapiro}}, \bibinfo {author}
  {\bibfnamefont {T.}~\bibnamefont {{Andruni\'{o}w}}}, \bibinfo {author}
  {\bibfnamefont {A.}~\bibnamefont {Dreuw}}, \ and\ \bibinfo {author}
  {\bibfnamefont {M.}~\bibnamefont {Olivucci}},\ }\href@noop {} {\bibfield
  {journal} {\bibinfo  {journal} {J. Chem. Theory Comput.}\ }\textbf {\bibinfo
  {volume} {11}},\ \bibinfo {pages} {5758} (\bibinfo {year}
  {2015})}\BibitemShut {NoStop}%
\bibitem [{\citenamefont {Epifanovsky}\ and\ \citenamefont
  {Krylov}(2007)}]{Epifanovsky2007MP}%
  \BibitemOpen
  \bibfield  {author} {\bibinfo {author} {\bibfnamefont {E.}~\bibnamefont
  {Epifanovsky}}\ and\ \bibinfo {author} {\bibfnamefont {A.~I.}\ \bibnamefont
  {Krylov}},\ }\href@noop {} {\bibfield  {journal} {\bibinfo  {journal} {Mol.
  Phys.}\ }\textbf {\bibinfo {volume} {105}},\ \bibinfo {pages} {2515}
  (\bibinfo {year} {2007})}\BibitemShut {NoStop}%
\bibitem [{\citenamefont {Shepler}\ \emph {et~al.}(2008)\citenamefont
  {Shepler}, \citenamefont {Epifanovsky}, \citenamefont {Zhang}, \citenamefont
  {Bowman}, \citenamefont {Krylov},\ and\ \citenamefont
  {Morokuma}}]{Shepler2008JPCA}%
  \BibitemOpen
  \bibfield  {author} {\bibinfo {author} {\bibfnamefont {B.~C.}\ \bibnamefont
  {Shepler}}, \bibinfo {author} {\bibfnamefont {E.}~\bibnamefont
  {Epifanovsky}}, \bibinfo {author} {\bibfnamefont {P.}~\bibnamefont {Zhang}},
  \bibinfo {author} {\bibfnamefont {J.~M.}\ \bibnamefont {Bowman}}, \bibinfo
  {author} {\bibfnamefont {A.~I.}\ \bibnamefont {Krylov}}, \ and\ \bibinfo
  {author} {\bibfnamefont {K.}~\bibnamefont {Morokuma}},\ }\href@noop {}
  {\bibfield  {journal} {\bibinfo  {journal} {J. Phys. Chem. A}\ }\textbf
  {\bibinfo {volume} {112}},\ \bibinfo {pages} {13267} (\bibinfo {year}
  {2008})}\BibitemShut {NoStop}%
\bibitem [{\citenamefont {Minezawa}\ and\ \citenamefont
  {Gordon}(2009)}]{Minezawa2009JPCA}%
  \BibitemOpen
  \bibfield  {author} {\bibinfo {author} {\bibfnamefont {N.}~\bibnamefont
  {Minezawa}}\ and\ \bibinfo {author} {\bibfnamefont {M.~S.}\ \bibnamefont
  {Gordon}},\ }\href@noop {} {\bibfield  {journal} {\bibinfo  {journal} {J.
  Phys. Chem. A}\ }\textbf {\bibinfo {volume} {113}},\ \bibinfo {pages} {12749}
  (\bibinfo {year} {2009})}\BibitemShut {NoStop}%
\bibitem [{\citenamefont {Andersson}\ \emph {et~al.}(1990)\citenamefont
  {Andersson}, \citenamefont {Malmqvist}, \citenamefont {Roos}, \citenamefont
  {Sadlej},\ and\ \citenamefont {Wolinski}}]{Andersson1990JPC}%
  \BibitemOpen
  \bibfield  {author} {\bibinfo {author} {\bibfnamefont {K.}~\bibnamefont
  {Andersson}}, \bibinfo {author} {\bibfnamefont {P.-{\AA}.}\ \bibnamefont
  {Malmqvist}}, \bibinfo {author} {\bibfnamefont {B.~O.}\ \bibnamefont {Roos}},
  \bibinfo {author} {\bibfnamefont {A.~J.}\ \bibnamefont {Sadlej}}, \ and\
  \bibinfo {author} {\bibfnamefont {K.}~\bibnamefont {Wolinski}},\ }\href@noop
  {} {\bibfield  {journal} {\bibinfo  {journal} {{J. Phys. Chem.}}\ }\textbf
  {\bibinfo {volume} {94}},\ \bibinfo {pages} {5483} (\bibinfo {year}
  {1990})}\BibitemShut {NoStop}%
\bibitem [{\citenamefont {Andersson}\ \emph {et~al.}(1992)\citenamefont
  {Andersson}, \citenamefont {Malmqvist},\ and\ \citenamefont
  {Roos}}]{Andersson1992JCP}%
  \BibitemOpen
  \bibfield  {author} {\bibinfo {author} {\bibfnamefont {K.}~\bibnamefont
  {Andersson}}, \bibinfo {author} {\bibfnamefont {P.-{\AA}.}\ \bibnamefont
  {Malmqvist}}, \ and\ \bibinfo {author} {\bibfnamefont {B.~O.}\ \bibnamefont
  {Roos}},\ }\href@noop {} {\bibfield  {journal} {\bibinfo  {journal} {{J.
  Chem. Phys.}}\ }\textbf {\bibinfo {volume} {96}},\ \bibinfo {pages} {1218}
  (\bibinfo {year} {1992})}\BibitemShut {NoStop}%
\bibitem [{\citenamefont {Finley}\ \emph {et~al.}(1998)\citenamefont {Finley},
  \citenamefont {Malmqvist}, \citenamefont {Roos},\ and\ \citenamefont
  {Serrano-Andr{\'{e}s}}}]{Finley1998CPL}%
  \BibitemOpen
  \bibfield  {author} {\bibinfo {author} {\bibfnamefont {J.}~\bibnamefont
  {Finley}}, \bibinfo {author} {\bibfnamefont {P.-{\AA}.}\ \bibnamefont
  {Malmqvist}}, \bibinfo {author} {\bibfnamefont {B.~O.}\ \bibnamefont {Roos}},
  \ and\ \bibinfo {author} {\bibfnamefont {L.}~\bibnamefont
  {Serrano-Andr{\'{e}s}}},\ }\href@noop {} {\bibfield  {journal} {\bibinfo
  {journal} {{Chem. Phys. Lett.}}\ }\textbf {\bibinfo {volume} {288}},\
  \bibinfo {pages} {299} (\bibinfo {year} {1998})}\BibitemShut {NoStop}%
\bibitem [{\citenamefont {Shiozaki}\ \emph {et~al.}(2011)\citenamefont
  {Shiozaki}, \citenamefont {{Gy\H{o}rffy}}, \citenamefont {Celani},\ and\
  \citenamefont {Werner}}]{Shiozaki2011JCP3}%
  \BibitemOpen
  \bibfield  {author} {\bibinfo {author} {\bibfnamefont {T.}~\bibnamefont
  {Shiozaki}}, \bibinfo {author} {\bibfnamefont {W.}~\bibnamefont
  {{Gy\H{o}rffy}}}, \bibinfo {author} {\bibfnamefont {P.}~\bibnamefont
  {Celani}}, \ and\ \bibinfo {author} {\bibfnamefont {H.-J.}\ \bibnamefont
  {Werner}},\ }\href@noop {} {\bibfield  {journal} {\bibinfo  {journal} {{J.
  Chem. Phys.}}\ }\textbf {\bibinfo {volume} {135}},\ \bibinfo {pages} {081106}
  (\bibinfo {year} {2011})}\BibitemShut {NoStop}%
\bibitem [{\citenamefont {{Serrano-Andr\'{e}s}}\ \emph
  {et~al.}(2005)\citenamefont {{Serrano-Andr\'{e}s}}, \citenamefont
  {{Merch\'an}},\ and\ \citenamefont {Lindh}}]{SerranoAndres2005JCP}%
  \BibitemOpen
  \bibfield  {author} {\bibinfo {author} {\bibfnamefont {L.}~\bibnamefont
  {{Serrano-Andr\'{e}s}}}, \bibinfo {author} {\bibfnamefont {M.}~\bibnamefont
  {{Merch\'an}}}, \ and\ \bibinfo {author} {\bibfnamefont {R.}~\bibnamefont
  {Lindh}},\ }\href@noop {} {\bibfield  {journal} {\bibinfo  {journal} {{J.
  Chem. Phys.}}\ }\textbf {\bibinfo {volume} {122}},\ \bibinfo {pages} {104107}
  (\bibinfo {year} {2005})}\BibitemShut {NoStop}%
\bibitem [{\citenamefont {Granovsky}(2011)}]{Granovsky2011JCP}%
  \BibitemOpen
  \bibfield  {author} {\bibinfo {author} {\bibfnamefont {A.~A.}\ \bibnamefont
  {Granovsky}},\ }\href@noop {} {\bibfield  {journal} {\bibinfo  {journal} {{J.
  Chem. Phys.}}\ }\textbf {\bibinfo {volume} {134}},\ \bibinfo {pages} {214113}
  (\bibinfo {year} {2011})}\BibitemShut {NoStop}%
\bibitem [{\citenamefont {MacLeod}\ and\ \citenamefont
  {Shiozaki}(2015)}]{MacLeod2015JCP}%
  \BibitemOpen
  \bibfield  {author} {\bibinfo {author} {\bibfnamefont {M.~K.}\ \bibnamefont
  {MacLeod}}\ and\ \bibinfo {author} {\bibfnamefont {T.}~\bibnamefont
  {Shiozaki}},\ }\href@noop {} {\bibfield  {journal} {\bibinfo  {journal} {J.
  Chem. Phys.}\ }\textbf {\bibinfo {volume} {142}},\ \bibinfo {pages} {051103}
  (\bibinfo {year} {2015})}\BibitemShut {NoStop}%
\bibitem [{\citenamefont {Vlaisavljevich}\ and\ \citenamefont
  {Shiozaki}(2016)}]{Vlaisavljevich2016JCTC}%
  \BibitemOpen
  \bibfield  {author} {\bibinfo {author} {\bibfnamefont {B.}~\bibnamefont
  {Vlaisavljevich}}\ and\ \bibinfo {author} {\bibfnamefont {T.}~\bibnamefont
  {Shiozaki}},\ }\href@noop {} {\bibfield  {journal} {\bibinfo  {journal} {J.
  Chem. Theory Comput.}\ }\textbf {\bibinfo {volume} {12}},\ \bibinfo {pages}
  {3781} (\bibinfo {year} {2016})}\BibitemShut {NoStop}%
\bibitem [{\citenamefont {Celani}\ and\ \citenamefont
  {Werner}(2003)}]{Celani2003JCP}%
  \BibitemOpen
  \bibfield  {author} {\bibinfo {author} {\bibfnamefont {P.}~\bibnamefont
  {Celani}}\ and\ \bibinfo {author} {\bibfnamefont {H.-J.}\ \bibnamefont
  {Werner}},\ }\href@noop {} {\bibfield  {journal} {\bibinfo  {journal} {{J.
  Chem. Phys.}}\ }\textbf {\bibinfo {volume} {119}},\ \bibinfo {pages} {5044}
  (\bibinfo {year} {2003})}\BibitemShut {NoStop}%
\bibitem [{\citenamefont {Werner}\ \emph {et~al.}(2011)\citenamefont {Werner},
  \citenamefont {Knowles}, \citenamefont {Knizia}, \citenamefont {Manby},\ and\
  \citenamefont {Sch{\"u}tz}}]{molpro}%
  \BibitemOpen
  \bibfield  {author} {\bibinfo {author} {\bibfnamefont {H.-J.}\ \bibnamefont
  {Werner}}, \bibinfo {author} {\bibfnamefont {P.~J.}\ \bibnamefont {Knowles}},
  \bibinfo {author} {\bibfnamefont {G.}~\bibnamefont {Knizia}}, \bibinfo
  {author} {\bibfnamefont {F.~R.}\ \bibnamefont {Manby}}, \ and\ \bibinfo
  {author} {\bibfnamefont {M.}~\bibnamefont {Sch{\"u}tz}},\ }\href@noop {}
  {\bibfield  {journal} {\bibinfo  {journal} {WIREs Comput. Mol. Sci.}\
  }\textbf {\bibinfo {volume} {2}},\ \bibinfo {pages} {242} (\bibinfo {year}
  {2011})}\BibitemShut {NoStop}%
\bibitem [{\citenamefont {Mori}\ and\ \citenamefont
  {Kato}(2009)}]{Mori2009CPL}%
  \BibitemOpen
  \bibfield  {author} {\bibinfo {author} {\bibfnamefont {T.}~\bibnamefont
  {Mori}}\ and\ \bibinfo {author} {\bibfnamefont {S.}~\bibnamefont {Kato}},\
  }\href@noop {} {\bibfield  {journal} {\bibinfo  {journal} {Chem. Phys.
  Lett.}\ }\textbf {\bibinfo {volume} {476}},\ \bibinfo {pages} {97} (\bibinfo
  {year} {2009})}\BibitemShut {NoStop}%
\bibitem [{\citenamefont {Roos}\ and\ \citenamefont
  {Andersson}(1995)}]{Roos1995CPL}%
  \BibitemOpen
  \bibfield  {author} {\bibinfo {author} {\bibfnamefont {B.~O.}\ \bibnamefont
  {Roos}}\ and\ \bibinfo {author} {\bibfnamefont {K.}~\bibnamefont
  {Andersson}},\ }\href@noop {} {\bibfield  {journal} {\bibinfo  {journal}
  {{Chem. Phys. Lett.}}\ }\textbf {\bibinfo {volume} {245}},\ \bibinfo {pages}
  {215} (\bibinfo {year} {1995})}\BibitemShut {NoStop}%
\bibitem [{\citenamefont {Fabiano}\ \emph {et~al.}(2008)\citenamefont
  {Fabiano}, \citenamefont {Keal},\ and\ \citenamefont
  {Thiel}}]{Fabiano2008CP}%
  \BibitemOpen
  \bibfield  {author} {\bibinfo {author} {\bibfnamefont {E.}~\bibnamefont
  {Fabiano}}, \bibinfo {author} {\bibfnamefont {T.~W.}\ \bibnamefont {Keal}}, \
  and\ \bibinfo {author} {\bibfnamefont {W.}~\bibnamefont {Thiel}},\
  }\href@noop {} {\bibfield  {journal} {\bibinfo  {journal} {Chem. Phys.}\
  }\textbf {\bibinfo {volume} {349}},\ \bibinfo {pages} {334} (\bibinfo {year}
  {2008})}\BibitemShut {NoStop}%
\bibitem [{\citenamefont {Koch}\ \emph {et~al.}(1990)\citenamefont {Koch},
  \citenamefont {Jensen}, \citenamefont {J{\o}rgensen}, \citenamefont
  {Helgaker}, \citenamefont {Scuseria},\ and\ \citenamefont {{Schaefer
  III}}}]{Koch1990JCP2}%
  \BibitemOpen
  \bibfield  {author} {\bibinfo {author} {\bibfnamefont {H.}~\bibnamefont
  {Koch}}, \bibinfo {author} {\bibfnamefont {H.~J.~A.}\ \bibnamefont {Jensen}},
  \bibinfo {author} {\bibfnamefont {P.}~\bibnamefont {J{\o}rgensen}}, \bibinfo
  {author} {\bibfnamefont {T.}~\bibnamefont {Helgaker}}, \bibinfo {author}
  {\bibfnamefont {G.~E.}\ \bibnamefont {Scuseria}}, \ and\ \bibinfo {author}
  {\bibfnamefont {H.~F.}\ \bibnamefont {{Schaefer III}}},\ }\href@noop {}
  {\bibfield  {journal} {\bibinfo  {journal} {J. Chem. Phys.}\ }\textbf
  {\bibinfo {volume} {92}},\ \bibinfo {pages} {4924} (\bibinfo {year}
  {1990})}\BibitemShut {NoStop}%
\bibitem [{\citenamefont {Handy}\ and\ \citenamefont
  {Schaefer}(1984)}]{Handy1984JCP}%
  \BibitemOpen
  \bibfield  {author} {\bibinfo {author} {\bibfnamefont {N.~C.}\ \bibnamefont
  {Handy}}\ and\ \bibinfo {author} {\bibfnamefont {H.~F.}\ \bibnamefont
  {Schaefer}},\ }\href@noop {} {\bibfield  {journal} {\bibinfo  {journal} {J.
  Chem. Phys.}\ }\textbf {\bibinfo {volume} {81}},\ \bibinfo {pages} {5031}
  (\bibinfo {year} {1984})}\BibitemShut {NoStop}%
\bibitem [{\citenamefont {{Gy\H orffy}}\ \emph {et~al.}(2013)\citenamefont
  {{Gy\H orffy}}, \citenamefont {Shiozaki}, \citenamefont {Knizia},\ and\
  \citenamefont {Werner}}]{Gyorffy2013JCP}%
  \BibitemOpen
  \bibfield  {author} {\bibinfo {author} {\bibfnamefont {W.}~\bibnamefont
  {{Gy\H orffy}}}, \bibinfo {author} {\bibfnamefont {T.}~\bibnamefont
  {Shiozaki}}, \bibinfo {author} {\bibfnamefont {G.}~\bibnamefont {Knizia}}, \
  and\ \bibinfo {author} {\bibfnamefont {H.-J.}\ \bibnamefont {Werner}},\
  }\href@noop {} {\bibfield  {journal} {\bibinfo  {journal} {J. Chem. Phys.}\
  }\textbf {\bibinfo {volume} {138}},\ \bibinfo {pages} {104104} (\bibinfo
  {year} {2013})}\BibitemShut {NoStop}%
\bibitem [{\citenamefont {Fatehi}\ and\ \citenamefont
  {Subotnik}(2012)}]{Fatehi2012JPCL}%
  \BibitemOpen
  \bibfield  {author} {\bibinfo {author} {\bibfnamefont {S.}~\bibnamefont
  {Fatehi}}\ and\ \bibinfo {author} {\bibfnamefont {J.~E.}\ \bibnamefont
  {Subotnik}},\ }\href@noop {} {\bibfield  {journal} {\bibinfo  {journal} {J.
  Phys. Chem. Lett.}\ }\textbf {\bibinfo {volume} {3}},\ \bibinfo {pages}
  {2039} (\bibinfo {year} {2012})}\BibitemShut {NoStop}%
\bibitem [{\citenamefont {Birkholz}\ and\ \citenamefont
  {Schlegel}(2016)}]{Birkholz2016TCA}%
  \BibitemOpen
  \bibfield  {author} {\bibinfo {author} {\bibfnamefont {A.~B.}\ \bibnamefont
  {Birkholz}}\ and\ \bibinfo {author} {\bibfnamefont {H.~B.}\ \bibnamefont
  {Schlegel}},\ }\href@noop {} {\bibfield  {journal} {\bibinfo  {journal}
  {Theor. Chem. Acc.}\ }\textbf {\bibinfo {volume} {135}},\ \bibinfo {pages}
  {84} (\bibinfo {year} {2016})}\BibitemShut {NoStop}%
\bibitem [{bag()}]{bagel}%
  \BibitemOpen
  \href@noop {} {}\bibinfo {note} {{\sc bagel}, Brilliantly Advanced General
  Electronic-structure Library. http://www.nubakery.org under the GNU General
  Public License.}\BibitemShut {Stop}%
\bibitem [{\citenamefont {Barbatti}\ \emph {et~al.}(2004)\citenamefont
  {Barbatti}, \citenamefont {Paier},\ and\ \citenamefont
  {Lischka}}]{Barbatti2004JCP}%
  \BibitemOpen
  \bibfield  {author} {\bibinfo {author} {\bibfnamefont {M.}~\bibnamefont
  {Barbatti}}, \bibinfo {author} {\bibfnamefont {J.}~\bibnamefont {Paier}}, \
  and\ \bibinfo {author} {\bibfnamefont {H.}~\bibnamefont {Lischka}},\
  }\href@noop {} {\bibfield  {journal} {\bibinfo  {journal} {J. Chem. Phys.}\
  }\textbf {\bibinfo {volume} {121}},\ \bibinfo {pages} {11614} (\bibinfo
  {year} {2004})}\BibitemShut {NoStop}%
\bibitem [{\citenamefont {Azizi}\ \emph {et~al.}(2006)\citenamefont {Azizi},
  \citenamefont {Roos},\ and\ \citenamefont {Veryazov}}]{Azizi2006PCCP}%
  \BibitemOpen
  \bibfield  {author} {\bibinfo {author} {\bibfnamefont {Z.}~\bibnamefont
  {Azizi}}, \bibinfo {author} {\bibfnamefont {B.~O.}\ \bibnamefont {Roos}}, \
  and\ \bibinfo {author} {\bibfnamefont {V.}~\bibnamefont {Veryazov}},\
  }\href@noop {} {\bibfield  {journal} {\bibinfo  {journal} {Phys. Chem. Chem.
  Phys.}\ }\textbf {\bibinfo {volume} {8}},\ \bibinfo {pages} {2727} (\bibinfo
  {year} {2006})}\BibitemShut {NoStop}%
\bibitem [{\citenamefont {Gozem}\ \emph {et~al.}(2012)\citenamefont {Gozem},
  \citenamefont {Huntress}, \citenamefont {Schapiro}, \citenamefont {Lindh},
  \citenamefont {Granovsky}, \citenamefont {Angeli},\ and\ \citenamefont
  {Olivucci}}]{Gozem2012JCTC}%
  \BibitemOpen
  \bibfield  {author} {\bibinfo {author} {\bibfnamefont {S.}~\bibnamefont
  {Gozem}}, \bibinfo {author} {\bibfnamefont {M.}~\bibnamefont {Huntress}},
  \bibinfo {author} {\bibfnamefont {I.}~\bibnamefont {Schapiro}}, \bibinfo
  {author} {\bibfnamefont {R.}~\bibnamefont {Lindh}}, \bibinfo {author}
  {\bibfnamefont {A.~A.}\ \bibnamefont {Granovsky}}, \bibinfo {author}
  {\bibfnamefont {C.}~\bibnamefont {Angeli}}, \ and\ \bibinfo {author}
  {\bibfnamefont {M.}~\bibnamefont {Olivucci}},\ }\href@noop {} {\bibfield
  {journal} {\bibinfo  {journal} {J. Chem. Theory Comput.}\ }\textbf {\bibinfo
  {volume} {8}},\ \bibinfo {pages} {4069} (\bibinfo {year} {2012})}\BibitemShut
  {NoStop}%
\bibitem [{\citenamefont {Gozem}\ \emph {et~al.}(2014)\citenamefont {Gozem},
  \citenamefont {Melaccio}, \citenamefont {Valentini}, \citenamefont {Filatov},
  \citenamefont {Huix-Rotllant}, \citenamefont {{Ferr\'{e}}}, \citenamefont
  {Frutos}, \citenamefont {Angeli}, \citenamefont {Krylov}, \citenamefont
  {Granovsky}, \citenamefont {Lindh}, ,\ and\ \citenamefont
  {Olivucci}}]{Gozem2014JCTC}%
  \BibitemOpen
  \bibfield  {author} {\bibinfo {author} {\bibfnamefont {S.}~\bibnamefont
  {Gozem}}, \bibinfo {author} {\bibfnamefont {F.}~\bibnamefont {Melaccio}},
  \bibinfo {author} {\bibfnamefont {A.}~\bibnamefont {Valentini}}, \bibinfo
  {author} {\bibfnamefont {M.}~\bibnamefont {Filatov}}, \bibinfo {author}
  {\bibfnamefont {M.}~\bibnamefont {Huix-Rotllant}}, \bibinfo {author}
  {\bibfnamefont {N.}~\bibnamefont {{Ferr\'{e}}}}, \bibinfo {author}
  {\bibfnamefont {L.~M.}\ \bibnamefont {Frutos}}, \bibinfo {author}
  {\bibfnamefont {C.}~\bibnamefont {Angeli}}, \bibinfo {author} {\bibfnamefont
  {A.~I.}\ \bibnamefont {Krylov}}, \bibinfo {author} {\bibfnamefont {A.~A.}\
  \bibnamefont {Granovsky}}, \bibinfo {author} {\bibfnamefont {R.}~\bibnamefont
  {Lindh}}, , \ and\ \bibinfo {author} {\bibfnamefont {M.}~\bibnamefont
  {Olivucci}},\ }\href@noop {} {\bibfield  {journal} {\bibinfo  {journal} {J.
  Chem. Theory Comput.}\ }\textbf {\bibinfo {volume} {10}},\ \bibinfo {pages}
  {3074} (\bibinfo {year} {2014})}\BibitemShut {NoStop}%
\bibitem [{\citenamefont {Quenneville}\ and\ \citenamefont
  {{Mart\'{i}nez}}(2003)}]{Quenneville2003JPCA}%
  \BibitemOpen
  \bibfield  {author} {\bibinfo {author} {\bibfnamefont {J.}~\bibnamefont
  {Quenneville}}\ and\ \bibinfo {author} {\bibfnamefont {T.~J.}\ \bibnamefont
  {{Mart\'{i}nez}}},\ }\href@noop {} {\bibfield  {journal} {\bibinfo  {journal}
  {J. Phys. Chem. A}\ }\textbf {\bibinfo {volume} {107}},\ \bibinfo {pages}
  {829} (\bibinfo {year} {2003})}\BibitemShut {NoStop}%
\bibitem [{\citenamefont {Minezawa}\ and\ \citenamefont
  {Gordon}(2011)}]{Minezawa2011JPCA}%
  \BibitemOpen
  \bibfield  {author} {\bibinfo {author} {\bibfnamefont {N.}~\bibnamefont
  {Minezawa}}\ and\ \bibinfo {author} {\bibfnamefont {M.~S.}\ \bibnamefont
  {Gordon}},\ }\href@noop {} {\bibfield  {journal} {\bibinfo  {journal} {J.
  Phys. Chem. A.}\ }\textbf {\bibinfo {volume} {115}},\ \bibinfo {pages} {7901}
  (\bibinfo {year} {2011})}\BibitemShut {NoStop}%
\bibitem [{\citenamefont {Ioffe}\ and\ \citenamefont
  {Granovsky}(2013)}]{Ioffe2013JCTC}%
  \BibitemOpen
  \bibfield  {author} {\bibinfo {author} {\bibfnamefont {I.~N.}\ \bibnamefont
  {Ioffe}}\ and\ \bibinfo {author} {\bibfnamefont {A.~A.}\ \bibnamefont
  {Granovsky}},\ }\href@noop {} {\bibfield  {journal} {\bibinfo  {journal} {J.
  Chem. Theory Comput.}\ }\textbf {\bibinfo {volume} {9}},\ \bibinfo {pages}
  {4973} (\bibinfo {year} {2013})}\BibitemShut {NoStop}%
\bibitem [{\citenamefont {Lei}\ \emph {et~al.}(2014)\citenamefont {Lei},
  \citenamefont {Yu}, \citenamefont {Zhou}, \citenamefont {Zhu}, \citenamefont
  {Wen},\ and\ \citenamefont {Lin}}]{Lei2014JPCA}%
  \BibitemOpen
  \bibfield  {author} {\bibinfo {author} {\bibfnamefont {Y.}~\bibnamefont
  {Lei}}, \bibinfo {author} {\bibfnamefont {L.}~\bibnamefont {Yu}}, \bibinfo
  {author} {\bibfnamefont {B.}~\bibnamefont {Zhou}}, \bibinfo {author}
  {\bibfnamefont {C.}~\bibnamefont {Zhu}}, \bibinfo {author} {\bibfnamefont
  {Z.}~\bibnamefont {Wen}}, \ and\ \bibinfo {author} {\bibfnamefont {S.~H.}\
  \bibnamefont {Lin}},\ }\href@noop {} {\bibfield  {journal} {\bibinfo
  {journal} {J. Phys. Chem. A}\ }\textbf {\bibinfo {volume} {118}},\ \bibinfo
  {pages} {9021} (\bibinfo {year} {2014})}\BibitemShut {NoStop}%
\bibitem [{\citenamefont {Harabuchi}\ \emph {et~al.}(2014)\citenamefont
  {Harabuchi}, \citenamefont {Keipert}, \citenamefont {Zahariev}, \citenamefont
  {Taketsugu},\ and\ \citenamefont {Gordon}}]{Harabuchi2014JPCA}%
  \BibitemOpen
  \bibfield  {author} {\bibinfo {author} {\bibfnamefont {Y.}~\bibnamefont
  {Harabuchi}}, \bibinfo {author} {\bibfnamefont {K.}~\bibnamefont {Keipert}},
  \bibinfo {author} {\bibfnamefont {F.}~\bibnamefont {Zahariev}}, \bibinfo
  {author} {\bibfnamefont {T.}~\bibnamefont {Taketsugu}}, \ and\ \bibinfo
  {author} {\bibfnamefont {M.~S.}\ \bibnamefont {Gordon}},\ }\href@noop {}
  {\bibfield  {journal} {\bibinfo  {journal} {J. Phys. Chem. A}\ }\textbf
  {\bibinfo {volume} {118}},\ \bibinfo {pages} {11987} (\bibinfo {year}
  {2014})}\BibitemShut {NoStop}%
\bibitem [{\citenamefont {Waldeck}(1991)}]{Waldeck1991CR}%
  \BibitemOpen
  \bibfield  {author} {\bibinfo {author} {\bibfnamefont {D.~H.}\ \bibnamefont
  {Waldeck}},\ }\href@noop {} {\bibfield  {journal} {\bibinfo  {journal} {Chem.
  Rev.}\ }\textbf {\bibinfo {volume} {91}},\ \bibinfo {pages} {415} (\bibinfo
  {year} {1991})}\BibitemShut {NoStop}%
\bibitem [{\citenamefont {Tsien}(1998)}]{Tsien1998ARBC}%
  \BibitemOpen
  \bibfield  {author} {\bibinfo {author} {\bibfnamefont {R.~Y.}\ \bibnamefont
  {Tsien}},\ }\href@noop {} {\bibfield  {journal} {\bibinfo  {journal} {Annu.
  Rev. Biochem.}\ }\textbf {\bibinfo {volume} {67}},\ \bibinfo {pages} {509}
  (\bibinfo {year} {1998})}\BibitemShut {NoStop}%
\bibitem [{\citenamefont {Weber}\ \emph {et~al.}(1999)\citenamefont {Weber},
  \citenamefont {Helms}, \citenamefont {McCammon},\ and\ \citenamefont
  {Langhoff}}]{Weber1999PNAS}%
  \BibitemOpen
  \bibfield  {author} {\bibinfo {author} {\bibfnamefont {W.}~\bibnamefont
  {Weber}}, \bibinfo {author} {\bibfnamefont {V.}~\bibnamefont {Helms}},
  \bibinfo {author} {\bibfnamefont {A.~J.}\ \bibnamefont {McCammon}}, \ and\
  \bibinfo {author} {\bibfnamefont {P.~W.}\ \bibnamefont {Langhoff}},\
  }\href@noop {} {\bibfield  {journal} {\bibinfo  {journal} {Proc. Natl. Acad.
  Sci. U.S.A.}\ }\textbf {\bibinfo {volume} {96}},\ \bibinfo {pages} {6177}
  (\bibinfo {year} {1999})}\BibitemShut {NoStop}%
\bibitem [{\citenamefont {Martin}\ \emph {et~al.}(2004)\citenamefont {Martin},
  \citenamefont {Negri},\ and\ \citenamefont {Olivucci}}]{Martin2004JACS}%
  \BibitemOpen
  \bibfield  {author} {\bibinfo {author} {\bibfnamefont {M.~E.}\ \bibnamefont
  {Martin}}, \bibinfo {author} {\bibfnamefont {F.}~\bibnamefont {Negri}}, \
  and\ \bibinfo {author} {\bibfnamefont {M.}~\bibnamefont {Olivucci}},\
  }\href@noop {} {\bibfield  {journal} {\bibinfo  {journal} {J. Am. Chem.
  Soc.}\ }\textbf {\bibinfo {volume} {126}},\ \bibinfo {pages} {5452} (\bibinfo
  {year} {2004})}\BibitemShut {NoStop}%
\bibitem [{\citenamefont {Olsen}\ and\ \citenamefont
  {Smith}(2008)}]{Olsen2008JACS}%
  \BibitemOpen
  \bibfield  {author} {\bibinfo {author} {\bibfnamefont {S.}~\bibnamefont
  {Olsen}}\ and\ \bibinfo {author} {\bibfnamefont {S.~C.}\ \bibnamefont
  {Smith}},\ }\href@noop {} {\bibfield  {journal} {\bibinfo  {journal} {J. Am.
  Chem. Soc.}\ }\textbf {\bibinfo {volume} {130}},\ \bibinfo {pages} {8677}
  (\bibinfo {year} {2008})}\BibitemShut {NoStop}%
\bibitem [{\citenamefont {Park}\ and\ \citenamefont
  {Rhee}(2016)}]{Park2016JACS}%
  \BibitemOpen
  \bibfield  {author} {\bibinfo {author} {\bibfnamefont {J.~W.}\ \bibnamefont
  {Park}}\ and\ \bibinfo {author} {\bibfnamefont {Y.~M.}\ \bibnamefont
  {Rhee}},\ }\href@noop {} {\bibfield  {journal} {\bibinfo  {journal} {J. Am.
  Chem. Soc.}\ }\textbf {\bibinfo {volume} {138}},\ \bibinfo {pages} {13619}
  (\bibinfo {year} {2016})}\BibitemShut {NoStop}%
\bibitem [{\citenamefont {van Thor}(2009)}]{vanThor2009CSR}%
  \BibitemOpen
  \bibfield  {author} {\bibinfo {author} {\bibfnamefont {J.~J.}\ \bibnamefont
  {van Thor}},\ }\href@noop {} {\bibfield  {journal} {\bibinfo  {journal}
  {Chem. Soc. Rev.}\ }\textbf {\bibinfo {volume} {38}},\ \bibinfo {pages}
  {2935} (\bibinfo {year} {2009})}\BibitemShut {NoStop}%
\bibitem [{\citenamefont {Polyakov}\ \emph {et~al.}(2010)\citenamefont
  {Polyakov}, \citenamefont {Grigorenko}, \citenamefont {Epifanovsky},
  \citenamefont {Krylov},\ and\ \citenamefont {Nemukhin}}]{Polyakov2010JCTC}%
  \BibitemOpen
  \bibfield  {author} {\bibinfo {author} {\bibfnamefont {I.~V.}\ \bibnamefont
  {Polyakov}}, \bibinfo {author} {\bibfnamefont {B.~L.}\ \bibnamefont
  {Grigorenko}}, \bibinfo {author} {\bibfnamefont {E.~M.}\ \bibnamefont
  {Epifanovsky}}, \bibinfo {author} {\bibfnamefont {A.~I.}\ \bibnamefont
  {Krylov}}, \ and\ \bibinfo {author} {\bibfnamefont {A.~V.}\ \bibnamefont
  {Nemukhin}},\ }\href@noop {} {\bibfield  {journal} {\bibinfo  {journal} {J.
  Chem. Theory Comput.}\ }\textbf {\bibinfo {volume} {6}},\ \bibinfo {pages}
  {2377} (\bibinfo {year} {2010})}\BibitemShut {NoStop}%
\bibitem [{\citenamefont {Acharya}\ \emph {et~al.}(2017)\citenamefont
  {Acharya}, \citenamefont {Bogdanov}, \citenamefont {Grigorenko},
  \citenamefont {Bravaya}, \citenamefont {Nemukhin}, \citenamefont {Lukyanov},\
  and\ \citenamefont {Krylov}}]{Acharya2017CR}%
  \BibitemOpen
  \bibfield  {author} {\bibinfo {author} {\bibfnamefont {A.}~\bibnamefont
  {Acharya}}, \bibinfo {author} {\bibfnamefont {A.~M.}\ \bibnamefont
  {Bogdanov}}, \bibinfo {author} {\bibfnamefont {B.~L.}\ \bibnamefont
  {Grigorenko}}, \bibinfo {author} {\bibfnamefont {K.~B.}\ \bibnamefont
  {Bravaya}}, \bibinfo {author} {\bibfnamefont {A.~V.}\ \bibnamefont
  {Nemukhin}}, \bibinfo {author} {\bibfnamefont {K.~A.}\ \bibnamefont
  {Lukyanov}}, \ and\ \bibinfo {author} {\bibfnamefont {A.~I.}\ \bibnamefont
  {Krylov}},\ }\href@noop {} {\bibfield  {journal} {\bibinfo  {journal} {Chem.
  Rev.}\ }\textbf {\bibinfo {volume} {117}},\ \bibinfo {pages} {758} (\bibinfo
  {year} {2017})}\BibitemShut {NoStop}%
\bibitem [{\citenamefont {Morozov}\ and\ \citenamefont
  {Groenhof}(2015)}]{Morozov2016ACIE}%
  \BibitemOpen
  \bibfield  {author} {\bibinfo {author} {\bibfnamefont {D.}~\bibnamefont
  {Morozov}}\ and\ \bibinfo {author} {\bibfnamefont {G.}~\bibnamefont
  {Groenhof}},\ }\href@noop {} {\bibfield  {journal} {\bibinfo  {journal}
  {Angew. Chem. Int. Ed.}\ }\textbf {\bibinfo {volume} {55}},\ \bibinfo {pages}
  {576} (\bibinfo {year} {2015})}\BibitemShut {NoStop}%
\bibitem [{\citenamefont {Minezawa}\ and\ \citenamefont
  {Gordon}(2012)}]{Minezawa2012JCP}%
  \BibitemOpen
  \bibfield  {author} {\bibinfo {author} {\bibfnamefont {N.}~\bibnamefont
  {Minezawa}}\ and\ \bibinfo {author} {\bibfnamefont {M.~S.}\ \bibnamefont
  {Gordon}},\ }\href@noop {} {\bibfield  {journal} {\bibinfo  {journal} {J.
  Chem. Phys.}\ }\textbf {\bibinfo {volume} {137}},\ \bibinfo {pages} {034116}
  (\bibinfo {year} {2012})}\BibitemShut {NoStop}%
\bibitem [{\citenamefont {Zhang}\ \emph {et~al.}(2014)\citenamefont {Zhang},
  \citenamefont {Chen}, \citenamefont {Cui}, \citenamefont {Fang},\ and\
  \citenamefont {Thiel}}]{Zhang2014ACIE}%
  \BibitemOpen
  \bibfield  {author} {\bibinfo {author} {\bibfnamefont {Q.}~\bibnamefont
  {Zhang}}, \bibinfo {author} {\bibfnamefont {X.}~\bibnamefont {Chen}},
  \bibinfo {author} {\bibfnamefont {G.}~\bibnamefont {Cui}}, \bibinfo {author}
  {\bibfnamefont {W.-H.}\ \bibnamefont {Fang}}, \ and\ \bibinfo {author}
  {\bibfnamefont {W.}~\bibnamefont {Thiel}},\ }\href@noop {} {\bibfield
  {journal} {\bibinfo  {journal} {Angew. Chem. Int. Ed.}\ }\textbf {\bibinfo
  {volume} {53}},\ \bibinfo {pages} {8649} (\bibinfo {year}
  {2014})}\BibitemShut {NoStop}%
\bibitem [{\citenamefont {Mandal}\ \emph {et~al.}(2004)\citenamefont {Mandal},
  \citenamefont {Tahara},\ and\ \citenamefont {Meech}}]{Mandal2004JPCB}%
  \BibitemOpen
  \bibfield  {author} {\bibinfo {author} {\bibfnamefont {D.}~\bibnamefont
  {Mandal}}, \bibinfo {author} {\bibfnamefont {T.}~\bibnamefont {Tahara}}, \
  and\ \bibinfo {author} {\bibfnamefont {S.~R.}\ \bibnamefont {Meech}},\
  }\href@noop {} {\bibfield  {journal} {\bibinfo  {journal} {J. Phys. Chem. B}\
  }\textbf {\bibinfo {volume} {108}},\ \bibinfo {pages} {1102} (\bibinfo {year}
  {2004})}\BibitemShut {NoStop}%
\bibitem [{\citenamefont {Olsen}(2010)}]{Olsen2010JCTC}%
  \BibitemOpen
  \bibfield  {author} {\bibinfo {author} {\bibfnamefont {S.}~\bibnamefont
  {Olsen}},\ }\href@noop {} {\bibfield  {journal} {\bibinfo  {journal} {J.
  Chem. Theory Comput.}\ }\textbf {\bibinfo {volume} {6}},\ \bibinfo {pages}
  {1089} (\bibinfo {year} {2010})}\BibitemShut {NoStop}%
\bibitem [{\citenamefont {Olsen}\ and\ \citenamefont
  {McKenzie}(2009)}]{Olsen2009JCP}%
  \BibitemOpen
  \bibfield  {author} {\bibinfo {author} {\bibfnamefont {S.}~\bibnamefont
  {Olsen}}\ and\ \bibinfo {author} {\bibfnamefont {R.~H.}\ \bibnamefont
  {McKenzie}},\ }\href@noop {} {\bibfield  {journal} {\bibinfo  {journal} {J.
  Chem. Phys.}\ }\textbf {\bibinfo {volume} {130}},\ \bibinfo {pages} {184302}
  (\bibinfo {year} {2009})}\BibitemShut {NoStop}%
\bibitem [{\citenamefont {Galloy}\ and\ \citenamefont
  {Lorquet}(1977)}]{Galloy1977JCP}%
  \BibitemOpen
  \bibfield  {author} {\bibinfo {author} {\bibfnamefont {C.}~\bibnamefont
  {Galloy}}\ and\ \bibinfo {author} {\bibfnamefont {J.~C.}\ \bibnamefont
  {Lorquet}},\ }\href@noop {} {\bibfield  {journal} {\bibinfo  {journal} {J.
  Chem. Phys.}\ }\textbf {\bibinfo {volume} {67}},\ \bibinfo {pages} {4672}
  (\bibinfo {year} {1977})}\BibitemShut {NoStop}%
\end{thebibliography}
\end{document}